\documentclass[%
 reprint,
superscriptaddress,
bibnotes,
 amsmath,amssymb,
 aps,
 prl,
]{revtex4-2}

\usepackage{graphicx}
\usepackage{dcolumn}
\usepackage{bm}
\usepackage[colorlinks=True,linkcolor=blue,anchorcolor=blue,citecolor=blue,urlcolor=blue,CJKbookmarks=true]{hyperref}

\begin{document}


\title{Topological Defects and Geometrical Frustrations in Fourier Photonic Simulator}






\affiliation{Zhejiang Key Laboratory of Micro-Nano Quantum Chips and Quantum Control, School of Physics, and State Key Laboratory for
Extreme Photonics and Instrumentation, Zhejiang University, Hangzhou 310027, China} 
\affiliation{College of Optical Science and Engineering, Zhejiang University, Hangzhou 310027, Zhejiang Province, China}
\affiliation{Hefei National Laboratory, Hefei 230088, China}
\affiliation{Institute for Advanced Study in Physics and School of Physics, Zhejiang University, Hangzhou, 310058, China}

\author{Yuxuan Sun$^{1,*}$}
\author{Weiru Fan$^{1,*,\dagger}$}
\author{Xingqi Xu$^{1}$}
\author{Da-Wei Wang$^{1,2,3,\ddagger}$}
\author{Hai-Qing Lin$^{4,\S}$}

\noaffiliation

\begin{abstract}
The XY models with continuous spin orientation play a pivotal role in understanding topological phase transitions and emergent frustration phenomena, such as superconducting and superfluid phase transitions. However, the complex energy landscapes arising from frustrated lattice geometries and competing spin interactions make these models computationally intractable. To address this challenge, we designed a programmable photonic spin simulator capable of emulating the XY models with tunable lattice geometries and spin couplings, allowing systematic exploration of their statistical properties. We experimentally observed the Berezinskii-Kosterlitz-Thouless transition in a square-lattice XY model with nearest-neighbor interactions and determined its critical temperature. Expanding to frustrated systems, we implemented the approach in triangular and honeycomb lattices, uncovering sophisticated phase transitions and frustration effects, which were consistent with theoretical predictions. This versatile platform opens avenues for probing unexplored XY model phenomena across diverse geometries and interaction schemes, with potential applications in solving complex optimization and machine learning problems.
\end{abstract}

\maketitle

The XY models \cite{kosterlitz_critical_1974} provide profound insight into complex dynamics in multidisciplines in condensed matter physics \cite{mallik_superfluid_2022,christodoulou_observation_2021}, machine learning \cite{chen_continuous_2003, chen_application_2014}, and continuous optimization \cite{andreasson_introduction_2020}. These models are characterized by continuous spin orientations in two-dimensional lattices, which host a topological phase transition at finite temperature, that is, the Berezinskii-Kosterlitz-Thouless (BKT) phase transition \cite{kosterlitz_ordering_1973}. Recently, it has been successfully applied to explain phase transitions in superconducting \cite{mallik_superfluid_2022} and superfluid \cite{christodoulou_observation_2021} films, which are more consistent with the experimental results than the Bardeen–Cooper–Schrieffer (BCS) theory. To explore physical phenomena from XY models, sampling spin configurations from the phase space according to the Boltzmann distribution is required \cite{pathria_statistical_2016,landau_guide_2021}. This sampling process is computationally daunting due to the exponential increase in the number of spin configurations with the number of spins \cite{cooper_computational_1990}. In addition, the energy landscape of the XY models is rugged arising from frustrated lattice geometries and various spin interactions, resulting in the extremely long relaxation of sampling procedure for generating effective spin configurations \cite{takeda_boltzmann_2017}. As a result, the investigation of XY models is resource intensive and challenging to expand to a large scale on computers.\par

To mitigate this difficulty, analog simulators have emerged as powerful alternatives to replace conventional electronic computers. Platforms such as optical parametric oscillators \cite{takeda_boltzmann_2017, honjo_100000-spin_2021}, superconducting quantum circuits \cite{johnson_quantum_2011}, spatial light modulators (SLMs) \cite{pierangeli_large-scale_2019,yamashita_low-rank_2023,fan_programmable_2023,fang_experimental_2021,leonetti_optical_2021,luo_wavelength-division_2023,ouyang_programmable_2024}, degenerate cavities \cite{nixon_observing_2013, gershenzon_exact_2020}, nanolasers \cite{parto_realizing_2020} and polariton condensates \cite{berloff_realizing_2017,kalinin_polaritonic_2020} are used to encode spins, allowing efficient sampling and optimization. Among these, the free-space light-based photonic simulators stand out for their accessibility, scalability, and compatibility to existing algorithms \cite{pierangeli_large-scale_2019,fan_programmable_2023,yamashita_low-rank_2023}, owing to the use of readily segmentable and accessible wavefront for encoding spins. Through this simulator, the phase transition of spin glass models can be experimentally observed by annealing samplings\cite{fang_experimental_2021,leonetti_optical_2021,luo_wavelength-division_2023}, and combinatorial optimization problems can be solved by searching the ground states of spin glass models \cite{pierangeli_large-scale_2019,yamashita_low-rank_2023}. \par

Despite its versatility, previous studies mostly focused on the simulation of Ising models with binary value spins. Recently, continuous spins on square lattice without frustration have also been investigated \cite{feng_Spin_2024, yuSpatialOpticalSimulator2024}. However, the encoding and sampling of spin configurations in XY models with programmable lattice geometries and spin interactions are still challenging. These continuous spin models are more general in thermodynamics and can be directly applied to practical continuous optimization problems \cite{chen_continuous_2003, chen_application_2014}, for instance, continuous graph coloring \cite{cavaliere_optimization_2021}, angle synchronization problems \cite{singer_angular_2011} and max $k$-cut \cite{PhysRevApplied.17.024063}. 
Toward this goal, a photonic simulator has been constructed with delicate but sophisticated design and algorithm \cite{ouyang_programmable_2024}, where the BKT transition was studied. However, its scalability in lattice size is limited and the generalizability to different lattice geometries has not been substantiated.

\begin{figure*}[hbt]
\centering
\includegraphics[width=0.65\linewidth]{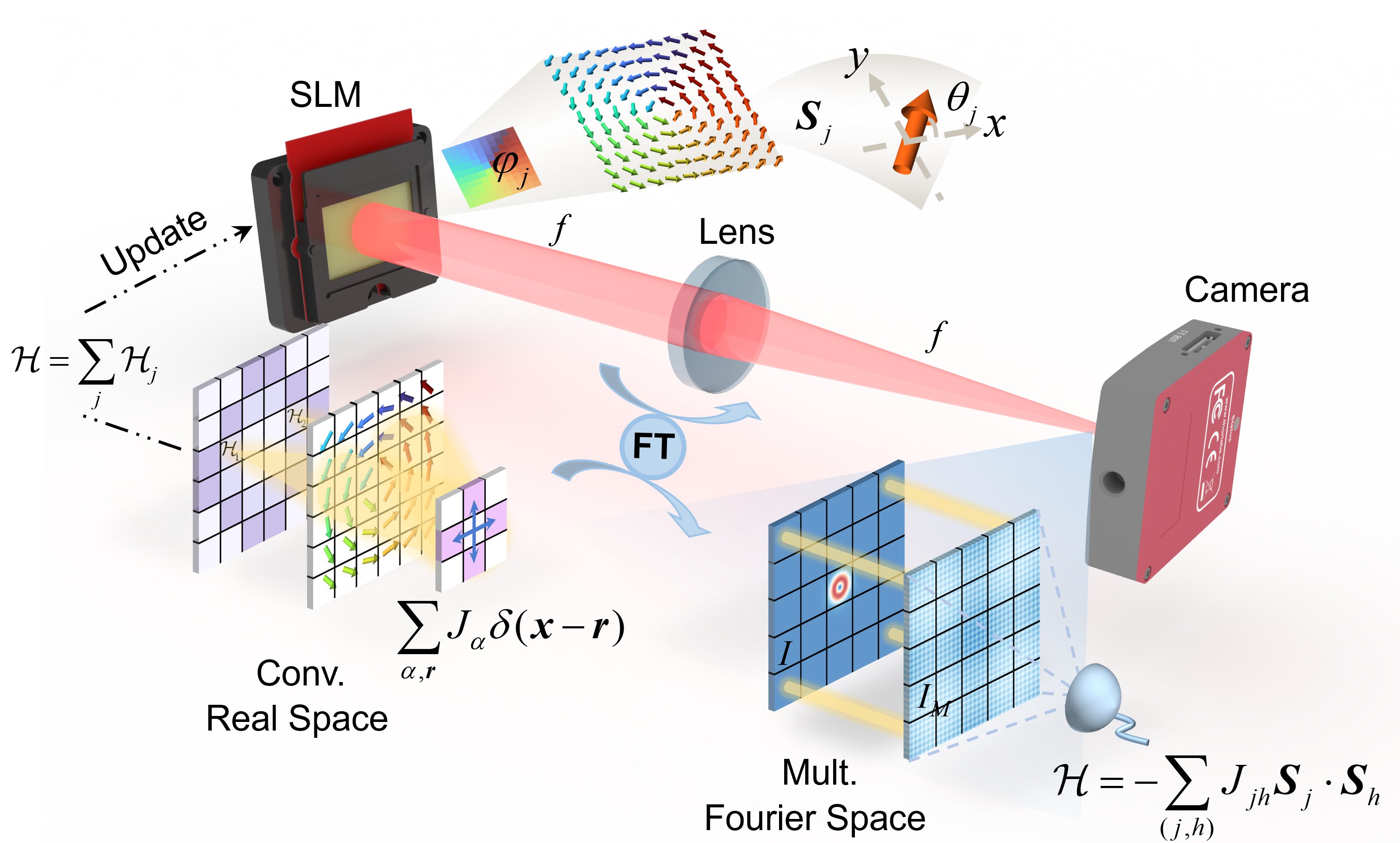}
\caption{\label{fig1}The principle of FPS. The XY spins $\boldsymbol{S}_j$ with angle $\theta_j$ are encoded to the wavefront phase $\varphi_j$ of a laser by an SLM. The modulated light passes through a lens, which performs an optical Fourier transform, and is finally detected by a camera on the back focal plane of the lens. A pointwise multiplication (Mult.) or Hadamard product is implemented between the light intensity $I$ and the Fourier mask $I_M$, which is equivalent to a convolution (Conv.) operation in the real space. Finally, the Hamiltonian of the XY model is obtained from the summation over all values of the products. Here, the Fourier mask has the effect of a filter for spin interactions, which can select the Hamiltonian of arbitrary range spin interactions. The MCMC algorithm is introduced to update the spin configurations by using the optically obtained Hamiltonian, thereby sampling spin configurations that obey the Boltzmann distribution. These samples can be employed to obtain important quantities of the XY models, such as the correlation function, helicity modulus and specific heat.}

\end{figure*}

In this Letter, we propose and implement a scalable and reconfigurable Fourier photonic simulator (FPS) to experimentally investigate the BKT phase transitions and study the frustration of XY models with programmable lattice geometries and spin interactions. By using an SLM that can simultaneously modulate the phase and amplitude of light, we encode spins into the phase of a light wavefront and engineer spin interactions via a reconfigurable Fourier mask \cite{fan_programmable_2023}.
To validate FPS, we constructed a ferromagnetic XY model on a square lattice, successfully observed the BKT phase transition, and determined the transition temperature. We further extend the approach to frustrated systems, including triangular and honeycomb lattices with competing antiferromagnetic spin interactions \cite{shaginyan_theoretical_2020}. The triangular lattice exhibits a dual BKT and Ising transition, while the honeycomb lattice hosts various ground states, depending on the interaction parameters. These findings highlight the interplay between lattice geometry, frustration, and topology and demonstrate the versatility of FPS.

\emph{Fourier-mask enabled photonic spin simulator}--The Hamiltonian of XY models is
\begin{align}
    \mathcal{H}&= -\sum_{(j,\,h)}J_{jh}\boldsymbol{S}_j\cdot\boldsymbol{S}_h\nonumber \\
&=-\sum_{\alpha\in \mathbb{Z}^+} J_\alpha\sum_{(j,\,h)_\alpha}\xi_j\xi_h\cos\left(\theta_j-\theta_h\right)~,
\label{equ1}
\end{align}
where $\boldsymbol{S}=\left(\cos\theta,~\sin\theta\right)$ with $\theta\in\left[0,~2\pi\right)$ is the spin of XY models \cite{kawamura_ordering_2011}, and $ J_{jh}$ denotes the interaction strength between the $j$-th and $h$-th spins. $J_{\alpha}$ is the interaction strength of the $\alpha$ range between neighbors (with $\alpha=1$ being the nearest neighbors). $\xi \in [-1,1]$ is a random variable which characterizes the disorder strength in {Mattis-type} spin glass models \cite{mattis_solvable_1976}. 
The $(j,h)_\alpha$ under the summation designates spin pairs, such as $(j,h)_1=\left\langle j,h\right\rangle$ for the nearest-neighbor (NN) spin pairs and $(j,h)_2=\left\langle\left\langle j,h\right\rangle\right\rangle$ for the next-nearest-neighbor (NNN) spin pairs. 

In the FPS, the SLM pixel arrays correspond to the spin lattices with the spin orientation angles $\theta$ encoded in the modulated phases of light $\varphi$, and the disorder strength $\xi$ encoded in the light field amplitude $A$ (see FIG.~\ref{fig1} and End Matter). The light reflected from the SLM is transformed by a lens to its Fourier plane, where we place a camera to measure the light intensity $I(\bm{u})$, with $\bm{u}$ being the coordinates on the camera. By specifically designing a mask $I_M(\bm{u})$ and make integration of $I(\bm{u})I_M(\bm{u})$ we can obtain the Hamiltonian of the spins with various XY interactions in Eq.~(\ref{equ1}) (see End Matter and the Supplementary Materials \cite{supp} for more details). 

\begin{figure*}[hbt]
\centering
\includegraphics[width=0.9\linewidth]{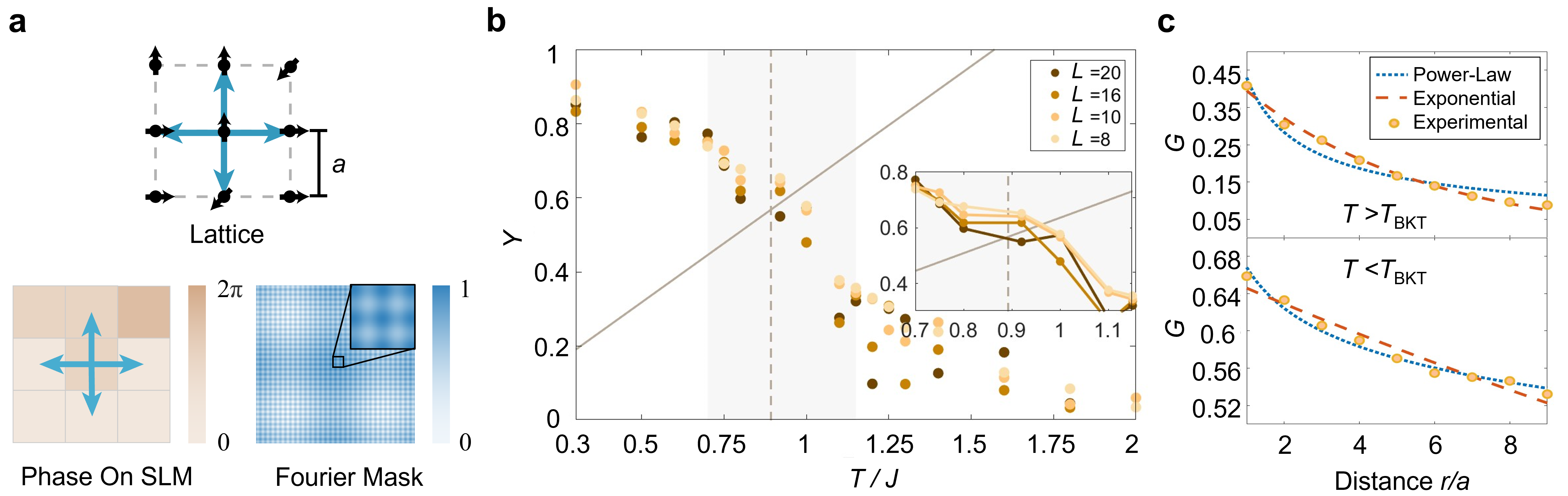}
\caption{\label{fig2} Observing the BKT transition of ferromagnetic XY model on the square lattice. (a) The lattice structure of a square lattice, the spin arrangement on the SLM, and the Fourier mask for NN interactions with periodic boundary conditions, respectively. $a$ is the lattice constant. The blue arrows represent spin coupling. (b) The helicity modulus ($Y$) as functions of temperature across different lattice sizes ($L\times L$=8$\times$8, 10$\times$10, 16$\times$16, 20$\times$20). The solid gray line represents $\tilde{Y} = \frac{2}{\pi}T$. The BKT transition is identified by the intersection of $Y(T)$ and $ \tilde{Y}(T)$. The dashed gray line indicates $T\approx0.8922$, which is the $T_\text{BKT}$ of infinite lattice. Insert provides a zoomed-in view near the BKT transition temperature $T_\text{BKT}$. (c) The correlation function $G(r)$ with respect to the distance $r$ between spins at $T = 1.2> T_\text{BKT}$ (top) and $T = 0.8<T_\text{BKT}$ (bottom). Above $T_\text{BKT}$, $G(r)$ fits the exponential decay model better with $R^2=0.9926$, while the power-law decay mode fits better below $T_\text{BKT}$ with $R^2=0.9780$.}
\end{figure*}
This strategy is compatible with the free arrangement of spins on the SLM, enabling special lattice configurations beyond square lattices. For instance, honeycomb and kagome lattices can be constructed by introducing vacancies and displacements of spins on the SLM. Since the SLM we used in the experiment can also modulate light intensity, we can introduce vacancy sites by setting the light intensity on those pixels to be zero. After obtaining the Hamiltonian, we use the Markov chain Monte Carlo (MCMC) sampling algorithm, e.g., Metropolis-Hastings \cite{chib_understanding_1995} and parallel tempering \cite{earl_parallel_2005}, to update the spin configurations and to study XY models across a range of lattice geometries. The sampling loop of FPS is shown in FIG.~\ref{fig1} (see \cite{supp} for algorithm details). 

\emph{Ferromagnetic XY model on the square lattice}--We first construct the ferromagnetic XY model on the square lattice with NN interaction, as shown in FIG.~\ref{fig2}(a). In this case, the $J_\alpha$ in Eq. (\ref{equ1}) is $J\delta(\alpha-1)$ where $J$ is the interaction strength constant. Unlike the Ising model, the XY model has no phase transition from spontaneous symmetry breaking \cite{mermin_absence_1966,hohenberg_existence_1967}, but possesses a BKT phase transition driven by interactions between topological defects \cite{zinn-justin_quantum_2021}. These topological defects featured by vortices have significantly different dynamics above and below a phase transition temperature $T_{\text{BKT}}/J\approx0.8922$ \cite{kosterlitz_long_1972,kosterlitz_ordering_1973}. Vortices with opposite charges, whose spins rotate clockwise and anti-clockwise, respectively, are tightly bound at low temperatures and dissociate above $T_{\text{BKT}}$ \cite{kosterlitz_nobel_2017}.

To study BKT phase transition and determine $T_{\text{BKT}}$, we introduce the helicity modulus, $Y(T)=\frac{1}{N} \left[\frac{1}{2} \langle\mathcal{H}\rangle-\beta\left(\langle I_x^2\rangle-\langle I_x\rangle^2 \right)\right]$, where $\beta=(k_BT)^{-1}$ is the inverse temperature with the Boltzmann constant $k_B$ (we set $k_B=1$ for simplicity), and $N$ is the number of spins. $I_x=-\sum_{\langle j,\,h\rangle_x} \sin(\theta_h-\theta_j)$ is the spin current along the $x$ direction, and $\langle\cdots\rangle_x$ denotes the NN spin pairs along the $x$ direction, $\langle I_x^2\rangle-\langle I_x\rangle^2$ is fluctuation of the spin current. Predicted by the renormalization group theory \cite{kosterlitz_ordering_1973}, the value of the helicity modulus with an infinite lattice size has a jump from zero to $\frac{2}{\pi}T_\text{BKT}$ at the BKT phase transition point. Consequently, the intersection of $Y(T)$ and $ \tilde{Y}(T)= \frac{2}{\pi}T $ can be used to determine the value of $T_\text{BKT}$. Especially, according to the scaling theory for finite-size systems, the BKT transition temperature increases when we decrease the lattice size \cite{weber_monte_1988}. To verify this phenomenon, we simulate XY models on square lattices of various sizes ($N=8\times8$, $10\times10$, $16\times16$, and $20\times20$) by FPS. As shown 
in FIG.~\ref{fig2}(b), the intersections converge to $T/J=0.8922$ from above when we increase the lattice size, consistent with the theoretical prediction.

The BKT phase transition can be characterized from another perspective. The correlation function $G( \boldsymbol{r})= \langle \boldsymbol{S}(\boldsymbol{x}_j) \cdot \boldsymbol{S} (\boldsymbol{x}_j+\boldsymbol{r}) \rangle$ scales differently above and below the $T_\text{BKT}$ \cite{situ_dynamics_2020}. When $T>T_\text{BKT}$, $G( \boldsymbol{r})$ decays exponentially because the  vortices are randomly distributed. When $T<T_\text{BKT}$, vortex pairs with opposite charges are tightly bound, such that the $G( \boldsymbol{r})$ decays in a power-law manner, indicating a quasi-long range order. In FIG.~\ref{fig2}(c), we show $G( \boldsymbol{r})$ for the two different topological phases at $T/J=1.2$ and $0.8$. The exponential decay function fits better for the high-temperature phase, while the power-law decay function fits better for the low-temperature phase. We also calculated the number of free vortices as a function of the temperature, as well as the change of the topological invariants (see \cite{supp}), which are coherent with theoretical predictions. Therefore, FPS can be used to precisely simulate XY models and reveal the complex BKT phase transition.

\emph{Antiferromagnetic XY model on the triangular lattice}--Antiferromagnetic interactions can induce frustration in XY models on triangular lattices \cite{lee_phase_1998}. Spin pairs of nearest neighbors tend to align antiparallel due to the antiferromagnetic coupling. However, spins at three vertices within each triangular unit cannot simultaneously align antiparallel to each other. It is termed geometrical frustration and leads to novel phases and orders \cite{bergman_order-by-disorder_2007}. For instance, some researches claim that this geometrical frustration in antiferromagnetic triangular lattice induces an additional second-order phase transition, and the critical temperature is slightly higher than the BKT transition temperature \cite{olsson_two_1995,lee_phase_1998}. Nevertheless, many results support the existence of only one BKT phase transition \cite{minnhagen_nonuniversal_1985,yosefin_phase_1985}. To experimentally validate this, we construct the antiferromagnetic XY model on a triangular lattice by FPS, and employ Metropolis-Hastings algorithm to investigate the frustration and phase transitions.

\begin{figure}[t]
\centering
\includegraphics[width=1\linewidth]{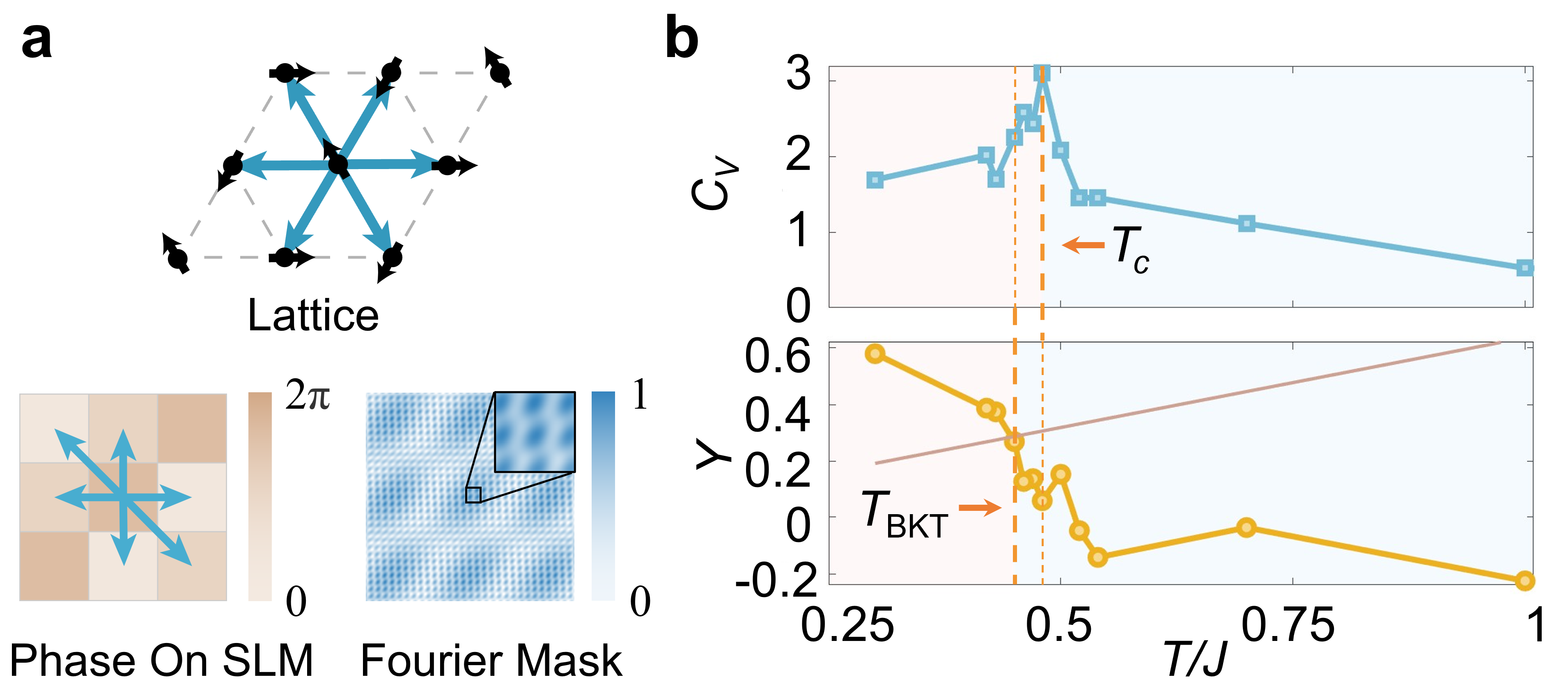}
\caption{\label{fig3}Simulating the antiferromagnetic XY model on a triangular lattice with $N = 12\times12$ spins. (a) The triangular lattice (upper panel), the interacting neighbors on the SLM (lower left), and the Fourier mask for NN interactions with periodic boundary conditions (lower right). (b) The specific heat ($C_V$) and helicity modulus ($Y$) as functions of temperature. $C_V(T)$ peaks at $T_c=0.48J$ indicating a second-order phase transition, while $Y(T)$ intersects $\tilde{Y}(T)=\frac{2}{\pi}T$ at $T_\text{BKT}\approx0.43J$, denoting a BKT transition with $T_\text{BKT}<T_c$.}
\end{figure}

{The triangular lattice is obtained on a square lattice by properly deforming and assigning interactions between each spin and its six neighbors. The corresponding deformed Fourier mask is shown in FIG.~\ref{fig3}(a) (see \cite{supp} for details).} We engineer the Fourier mask to include the NN interactions and the NNN interactions along the main diagonal of the square lattice. To distinguish the two different phase transitions, we use the specific heat $C_V = \frac{\beta^2}{N}\left(\langle \mathcal{H}^2\rangle-\langle \mathcal{H}\rangle^2 \right)$ to characterize the second-order phase transition \cite{lee_phase_1998}, while use the helicity modulus to characterize the BKT phase transition. As shown in FIG.~\ref{fig3}(b), the specific heat has a peak at $T_c=0.48J$ and the intersection of $Y(T)$ and $ \tilde{Y}(T)$ is at the $T_\text{BKT}\approx0.43J$. Our results reveal two distinct phase transitions for the antiferromagnetic XY model on a triangular lattice, in which the second-order phase transition occurs at a slightly higher temperature than the BKT phase transition, consistent with high-accuracy numerical simulations \cite{okumura_spin-chirality_2011}. Meanwhile, the geometrical frustration of the ground states is demonstrated by the angle between any two spin pairs, which is $2\pi/3$ within each triangular unit (see \cite{supp} for one of the ground states generated by FPS). Therefore, FPS is a reliable tool for studying complex phase-transition behavior involving frustrations.

\begin{figure*}[hbt]
\centering
\includegraphics[width=0.7\linewidth]{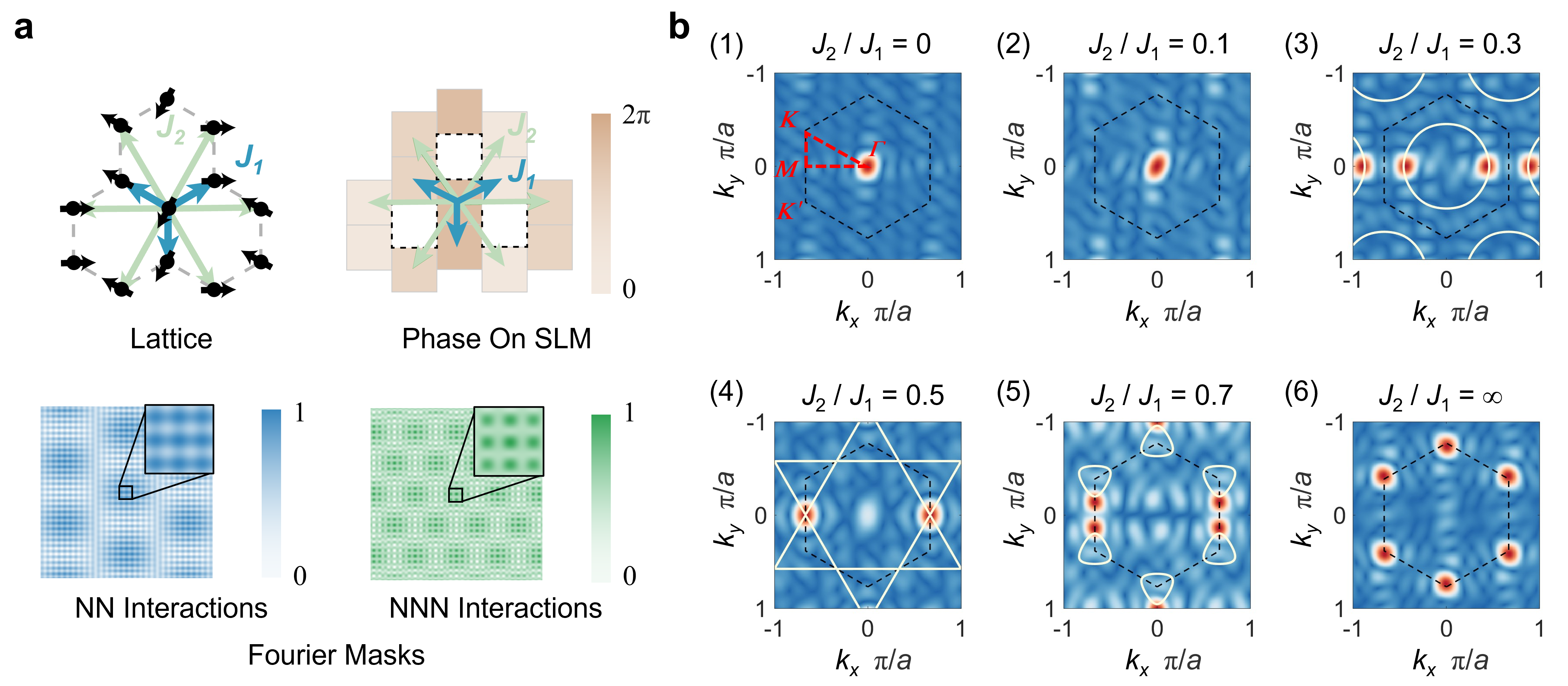}
\caption{\label{fig4}Searching the ground state of antiferromagnetic $J_1$-$J_2$ XY model on the honeycomb lattice with $N = 96$ spins. (a) The lattice structure of the honeycomb lattice (upper left), the spin arrangement on the SLM (upper right), and  the Fourier masks for NN interactions ($J_1$, lower left) and NNN interactions ($J_2$, lower right) with periodic boundary conditions. In building a honeycomb lattice from a square lattice, we set the light intensity in the dotted box areas in the SLM to be zero. (b) The ground states in reciprocal space with different $J_2/J_1$ ratios. The black dashed lines mark the boundary of the 1BZ, while the white solid lines indicate the theoretically predicted positions of the peaks \cite{di_ciolo_spiral_2014}.}
\end{figure*}

\emph{Antiferromagnetic $J_1$-$J_2$ XY model on the honeycomb lattice}--To demonstrate the ability of the FPS in lattice engineering, we further study the $J_1$-$J_2$ model on the honeycomb lattice. To construct a honeycomb lattice with the square lattice pixels of the SLM, we need to stretch the lattice, shift its sites and introduce vacancies, as shown in FIG.~\ref{fig4}(a). The NN and NNN spin interactions $J_1$ and $J_2$ are realized by Fourier masks and the ratio $J_2/J_1$ is tunable from zero to infinity \cite{di_ciolo_spiral_2014} by changing the superposition ratio of the two corresponding Fourier masks \cite{fan_programmable_2023}. The competition between the NN and NNN interactions leads to geometrical frustrations and highly degenerate ground states at the zero temperature. These can be characterized in reciprocal space by the distribution of spiral wave vectors within the first Brillouin zone (1BZ) \cite{di_ciolo_spiral_2014}.

We experimentally found ground states with different $J_2/J_1$ ratios, as shown in FIG.~\ref{fig4}(b). When $J_2=0$, only NN interaction is involved. In the ground state the nearest neighbor spin pairs are antiparallel. We obtain two sublattices with opposite spin orientations. Hence, the peak of the spiral wave vector is located at the center of the 1BZ ($\bm{\mathit{\Gamma}}$). It remains the case within the range $0\leq J_2/J_1\leq1/6$ (FIG.~\ref{fig4}(b1) and (b2)), indicating the N\'eel antiferromagnetic phase. When $1/6<J_2/J_1<1/2$, multiple peaks appear on a closed contour centered at the $\bm{\mathit{\Gamma}}$ point. The solid white lines in FIG.~\ref{fig4}(b3) show the theoretical predictions of all the degenerate ground states. In the experiment the system will randomly relax to some of the ground states, corresponding to the separated peaks on these curves. Specifically, the peaks for $J_2/J_1=1/2$ are along white lines passing through the $ \boldsymbol{M}$ points (FIG.~\ref{fig4}(b4)). When we further increase  the NNN interaction, $J_2/J_1>1/2$, the peaks appear on the closed contours around the $ \boldsymbol{K}$ and $ \boldsymbol{K}^\prime$ points (FIG.~\ref{fig4}(b5)). When $J_2/J_1\rightarrow\infty$, the orientations of NN spin pairs in each sublattice differ by an angle $2\pi/3$, and the peaks in the 1BZ are shifted to the $ \boldsymbol{K}$ and $ \boldsymbol{K}^\prime$ points, which is a $2\pi/3$ ordered phase (FIG.~\ref{fig4}(b6)). These experimental results are consistent with theoretical predictions \cite{di_ciolo_spiral_2014}, which demonstrates the ability of FPS for exploring spin systems with complex lattice geometries and spin interactions.

In conclusion, we present a programmable Fourier photonic spin simulator to solve the XY models with various lattice geometries and complex spin interactions. Such an FPS harnesses SLM to encode spins and the Fourier mask to obtain lattice geometries and spin interactions.  This approach enables a lattice-size-independent Hamiltonian calculation for the XY models. Using FPS, we experimentally demonstrated the BKT phase transition and spin frustration in ferromagnetic square lattices, antiferromagnetic triangular lattices, and antiferromagnetic $J_1$-$J_2$ honeycomb lattices. In these models, we precisely determined the critical transition temperatures, directly observed the geometrically frustrated states, and identified the degenerate ground-state configurations. These findings validate the universality of the FPS for exploring various phases and critical behaviors across the XY models, and can be extended to quantum spin systems \cite{maghrebi_continuous_2017,chen_equicalence_2018}. It also has the potential to solve continuous optimization problems, accelerate statistical learning algorithms, and establish {large-scale} optical neural networks. 

\emph{Acknowledgment}--This work was supported by the National Natural Science Foundation of China (Grant No. 12325412, 12088101, 12434020), National Key Research and Development Program of China (Grant No. 2024YFA1408900), Innovation Program for Quantum
Science and Technology (Grant No. 2021ZD0303200), and “Pioneer” and “Leading Goose” R\&D Program of Zhejiang (Grant No. 2025C01028) and the Fundamental Research Funds for the Central Universities.

\emph{Data availability}--Digital data associated with this work are available from Ref. \cite{supp}.


$\ $

$^{*}$Y.S. and W.F. contributed equally to this work

$^{\dagger}$Contact author: weiru\_fan@zju.edu.cn

$^{\ddagger}$Contact author: dwwang@zju.edu.cn

$^{\S}$Contact author: hqlin@zju.edu.cn


\begin{thebibliography}{61}%
\makeatletter
\providecommand \@ifxundefined [1]{%
 \@ifx{#1\undefined}
}%
\providecommand \@ifnum [1]{%
 \ifnum #1\expandafter \@firstoftwo
 \else \expandafter \@secondoftwo
 \fi
}%
\providecommand \@ifx [1]{%
 \ifx #1\expandafter \@firstoftwo
 \else \expandafter \@secondoftwo
 \fi
}%
\providecommand \natexlab [1]{#1}%
\providecommand \enquote  [1]{``#1''}%
\providecommand \bibnamefont  [1]{#1}%
\providecommand \bibfnamefont [1]{#1}%
\providecommand \citenamefont [1]{#1}%
\providecommand \href@noop [0]{\@secondoftwo}%
\providecommand \href [0]{\begingroup \@sanitize@url \@href}%
\providecommand \@href[1]{\@@startlink{#1}\@@href}%
\providecommand \@@href[1]{\endgroup#1\@@endlink}%
\providecommand \@sanitize@url [0]{\catcode `\\12\catcode `\$12\catcode `\&12\catcode `\#12\catcode `\^12\catcode `\_12\catcode `\%12\relax}%
\providecommand \@@startlink[1]{}%
\providecommand \@@endlink[0]{}%
\providecommand \url  [0]{\begingroup\@sanitize@url \@url }%
\providecommand \@url [1]{\endgroup\@href {#1}{\urlprefix }}%
\providecommand \urlprefix  [0]{URL }%
\providecommand \Eprint [0]{\href }%
\providecommand \doibase [0]{https://doi.org/}%
\providecommand \selectlanguage [0]{\@gobble}%
\providecommand \bibinfo  [0]{\@secondoftwo}%
\providecommand \bibfield  [0]{\@secondoftwo}%
\providecommand \translation [1]{[#1]}%
\providecommand \BibitemOpen [0]{}%
\providecommand \bibitemStop [0]{}%
\providecommand \bibitemNoStop [0]{.\EOS\space}%
\providecommand \EOS [0]{\spacefactor3000\relax}%
\providecommand \BibitemShut  [1]{\csname bibitem#1\endcsname}%
\let\auto@bib@innerbib\@empty
\bibitem [{\citenamefont {Kosterlitz}(1974)}]{kosterlitz_critical_1974}%
  \BibitemOpen
  \bibfield  {author} {\bibinfo {author} {\bibfnamefont {J.~M.}\ \bibnamefont {Kosterlitz}},\ }\bibfield  {title} {\bibinfo {title} {The critical properties of the two-dimensional {XY} model},\ }\href {https://doi.org/10.1088/0022-3719/7/6/005} {\bibfield  {journal} {\bibinfo  {journal} {J. Phys. C: Solid State Phys.}\ }\textbf {\bibinfo {volume} {7}},\ \bibinfo {pages} {1046} (\bibinfo {year} {1974})}\BibitemShut {NoStop}%
\bibitem [{\citenamefont {Mallik}\ \emph {et~al.}(2022)\citenamefont {Mallik}, \citenamefont {Ménard}, \citenamefont {Saïz}, \citenamefont {Witt}, \citenamefont {Lesueur}, \citenamefont {Gloter}, \citenamefont {Benfatto}, \citenamefont {Bibes},\ and\ \citenamefont {Bergeal}}]{mallik_superfluid_2022}%
  \BibitemOpen
  \bibfield  {author} {\bibinfo {author} {\bibfnamefont {S.}~\bibnamefont {Mallik}}, \bibinfo {author} {\bibfnamefont {G.~C.}\ \bibnamefont {Ménard}}, \bibinfo {author} {\bibfnamefont {G.}~\bibnamefont {Saïz}}, \bibinfo {author} {\bibfnamefont {H.}~\bibnamefont {Witt}}, \bibinfo {author} {\bibfnamefont {J.}~\bibnamefont {Lesueur}}, \bibinfo {author} {\bibfnamefont {A.}~\bibnamefont {Gloter}}, \bibinfo {author} {\bibfnamefont {L.}~\bibnamefont {Benfatto}}, \bibinfo {author} {\bibfnamefont {M.}~\bibnamefont {Bibes}},\ and\ \bibinfo {author} {\bibfnamefont {N.}~\bibnamefont {Bergeal}},\ }\bibfield  {title} {\bibinfo {title} {Superfluid stiffness of a {KTaO3}-based two-dimensional electron gas},\ }\href {https://doi.org/10.1038/s41467-022-32242-y} {\bibfield  {journal} {\bibinfo  {journal} {Nat. Commun.}\ }\textbf {\bibinfo {volume} {13}},\ \bibinfo {pages} {4625} (\bibinfo {year} {2022})}\BibitemShut {NoStop}%
\bibitem [{\citenamefont {Christodoulou}\ \emph {et~al.}(2021)\citenamefont {Christodoulou}, \citenamefont {Gałka}, \citenamefont {Dogra}, \citenamefont {Lopes}, \citenamefont {Schmitt},\ and\ \citenamefont {Hadzibabic}}]{christodoulou_observation_2021}%
  \BibitemOpen
  \bibfield  {author} {\bibinfo {author} {\bibfnamefont {P.}~\bibnamefont {Christodoulou}}, \bibinfo {author} {\bibfnamefont {M.}~\bibnamefont {Gałka}}, \bibinfo {author} {\bibfnamefont {N.}~\bibnamefont {Dogra}}, \bibinfo {author} {\bibfnamefont {R.}~\bibnamefont {Lopes}}, \bibinfo {author} {\bibfnamefont {J.}~\bibnamefont {Schmitt}},\ and\ \bibinfo {author} {\bibfnamefont {Z.}~\bibnamefont {Hadzibabic}},\ }\bibfield  {title} {\bibinfo {title} {Observation of first and second sound in a {BKT} superfluid},\ }\href {https://doi.org/10.1038/s41586-021-03537-9} {\bibfield  {journal} {\bibinfo  {journal} {Nature}\ }\textbf {\bibinfo {volume} {594}},\ \bibinfo {pages} {191} (\bibinfo {year} {2021})}\BibitemShut {NoStop}%
\bibitem [{\citenamefont {Chen}\ and\ \citenamefont {Murray}(2003)}]{chen_continuous_2003}%
  \BibitemOpen
  \bibfield  {author} {\bibinfo {author} {\bibfnamefont {H.}~\bibnamefont {Chen}}\ and\ \bibinfo {author} {\bibfnamefont {A.~F.}\ \bibnamefont {Murray}},\ }\bibfield  {title} {\bibinfo {title} {Continuous restricted {Boltzmann} machine with an implementable training algorithm},\ }\href {https://doi.org/10.1049/ip-vis:20030362} {\bibfield  {journal} {\bibinfo  {journal} {IEE P-Vis. Image Sign.}\ }\textbf {\bibinfo {volume} {150}},\ \bibinfo {pages} {153} (\bibinfo {year} {2003})}\BibitemShut {NoStop}%
\bibitem [{\citenamefont {Chen}\ \emph {et~al.}(2014)\citenamefont {Chen}, \citenamefont {Lu},\ and\ \citenamefont {Li}}]{chen_application_2014}%
  \BibitemOpen
  \bibfield  {author} {\bibinfo {author} {\bibfnamefont {Y.}~\bibnamefont {Chen}}, \bibinfo {author} {\bibfnamefont {L.}~\bibnamefont {Lu}},\ and\ \bibinfo {author} {\bibfnamefont {X.}~\bibnamefont {Li}},\ }\bibfield  {title} {\bibinfo {title} {Application of continuous restricted {Boltzmann} machine to identify multivariate geochemical anomaly},\ }\href {https://doi.org/10.1016/j.gexplo.2014.02.013} {\bibfield  {journal} {\bibinfo  {journal} {J. Geochem. Explor.}\ }\textbf {\bibinfo {volume} {140}},\ \bibinfo {pages} {56} (\bibinfo {year} {2014})}\BibitemShut {NoStop}%
\bibitem [{\citenamefont {Andreasson}\ \emph {et~al.}(2020)\citenamefont {Andreasson}, \citenamefont {Evgrafov},\ and\ \citenamefont {Patriksson}}]{andreasson_introduction_2020}%
  \BibitemOpen
  \bibfield  {author} {\bibinfo {author} {\bibfnamefont {N.}~\bibnamefont {Andreasson}}, \bibinfo {author} {\bibfnamefont {A.}~\bibnamefont {Evgrafov}},\ and\ \bibinfo {author} {\bibfnamefont {M.}~\bibnamefont {Patriksson}},\ }\href@noop {} {\emph {\bibinfo {title} {An {Introduction} to {Continuous} {Optimization}: {Foundations} and {Fundamental} {Algorithms}}}}\ (\bibinfo  {publisher} {Courier Dover Publications},\ \bibinfo {year} {2020})\BibitemShut {NoStop}%
\bibitem [{\citenamefont {Kosterlitz}\ and\ \citenamefont {Thouless}(1973)}]{kosterlitz_ordering_1973}%
  \BibitemOpen
  \bibfield  {author} {\bibinfo {author} {\bibfnamefont {J.~M.}\ \bibnamefont {Kosterlitz}}\ and\ \bibinfo {author} {\bibfnamefont {D.~J.}\ \bibnamefont {Thouless}},\ }\bibfield  {title} {\bibinfo {title} {Ordering, metastability and phase transitions in two-dimensional systems},\ }\href {https://doi.org/10.1088/0022-3719/6/7/010} {\bibfield  {journal} {\bibinfo  {journal} {J. Phys. C: Solid State Phys.}\ }\textbf {\bibinfo {volume} {6}},\ \bibinfo {pages} {1181} (\bibinfo {year} {1973})}\BibitemShut {NoStop}%
\bibitem [{\citenamefont {Pathria}(2016)}]{pathria_statistical_2016}%
  \BibitemOpen
  \bibfield  {author} {\bibinfo {author} {\bibfnamefont {R.~K.}\ \bibnamefont {Pathria}},\ }\href@noop {} {\emph {\bibinfo {title} {Statistical {Mechanics}}}}\ (\bibinfo  {publisher} {Elsevier},\ \bibinfo {year} {2016})\BibitemShut {NoStop}%
\bibitem [{\citenamefont {Landau}\ and\ \citenamefont {Binder}(2021)}]{landau_guide_2021}%
  \BibitemOpen
  \bibfield  {author} {\bibinfo {author} {\bibfnamefont {D.}~\bibnamefont {Landau}}\ and\ \bibinfo {author} {\bibfnamefont {K.}~\bibnamefont {Binder}},\ }\href@noop {} {\emph {\bibinfo {title} {A {Guide} to {Monte} {Carlo} {Simulations} in {Statistical} {Physics}}}}\ (\bibinfo  {publisher} {Cambridge University Press},\ \bibinfo {year} {2021})\BibitemShut {NoStop}%
\bibitem [{\citenamefont {Cooper}(1990)}]{cooper_computational_1990}%
  \BibitemOpen
  \bibfield  {author} {\bibinfo {author} {\bibfnamefont {G.~F.}\ \bibnamefont {Cooper}},\ }\bibfield  {title} {\bibinfo {title} {The computational complexity of probabilistic inference using bayesian belief networks},\ }\href {https://doi.org/10.1016/0004-3702(90)90060-D} {\bibfield  {journal} {\bibinfo  {journal} {Artif. Intell.}\ }\textbf {\bibinfo {volume} {42}},\ \bibinfo {pages} {393} (\bibinfo {year} {1990})}\BibitemShut {NoStop}%
\bibitem [{\citenamefont {Takeda}\ \emph {et~al.}(2017)\citenamefont {Takeda}, \citenamefont {Tamate}, \citenamefont {Yamamoto}, \citenamefont {Takesue}, \citenamefont {Inagaki},\ and\ \citenamefont {Utsunomiya}}]{takeda_boltzmann_2017}%
  \BibitemOpen
  \bibfield  {author} {\bibinfo {author} {\bibfnamefont {Y.}~\bibnamefont {Takeda}}, \bibinfo {author} {\bibfnamefont {S.}~\bibnamefont {Tamate}}, \bibinfo {author} {\bibfnamefont {Y.}~\bibnamefont {Yamamoto}}, \bibinfo {author} {\bibfnamefont {H.}~\bibnamefont {Takesue}}, \bibinfo {author} {\bibfnamefont {T.}~\bibnamefont {Inagaki}},\ and\ \bibinfo {author} {\bibfnamefont {S.}~\bibnamefont {Utsunomiya}},\ }\bibfield  {title} {\bibinfo {title} {Boltzmann sampling for an {XY} model using a non-degenerate optical parametric oscillator network},\ }\href {https://doi.org/10.1088/2058-9565/aa923b} {\bibfield  {journal} {\bibinfo  {journal} {Quantum Sci. Technol.}\ }\textbf {\bibinfo {volume} {3}},\ \bibinfo {pages} {014004} (\bibinfo {year} {2017})}\BibitemShut {NoStop}%
\bibitem [{\citenamefont {Honjo}\ \emph {et~al.}(2021)\citenamefont {Honjo}, \citenamefont {Sonobe}, \citenamefont {Inaba}, \citenamefont {Inagaki}, \citenamefont {Ikuta}, \citenamefont {Yamada}, \citenamefont {Kazama}, \citenamefont {Enbutsu}, \citenamefont {Umeki}, \citenamefont {Kasahara}, \citenamefont {Kawarabayashi},\ and\ \citenamefont {Takesue}}]{honjo_100000-spin_2021}%
  \BibitemOpen
  \bibfield  {author} {\bibinfo {author} {\bibfnamefont {T.}~\bibnamefont {Honjo}}, \bibinfo {author} {\bibfnamefont {T.}~\bibnamefont {Sonobe}}, \bibinfo {author} {\bibfnamefont {K.}~\bibnamefont {Inaba}}, \bibinfo {author} {\bibfnamefont {T.}~\bibnamefont {Inagaki}}, \bibinfo {author} {\bibfnamefont {T.}~\bibnamefont {Ikuta}}, \bibinfo {author} {\bibfnamefont {Y.}~\bibnamefont {Yamada}}, \bibinfo {author} {\bibfnamefont {T.}~\bibnamefont {Kazama}}, \bibinfo {author} {\bibfnamefont {K.}~\bibnamefont {Enbutsu}}, \bibinfo {author} {\bibfnamefont {T.}~\bibnamefont {Umeki}}, \bibinfo {author} {\bibfnamefont {R.}~\bibnamefont {Kasahara}}, \bibinfo {author} {\bibfnamefont {K.-i.}\ \bibnamefont {Kawarabayashi}},\ and\ \bibinfo {author} {\bibfnamefont {H.}~\bibnamefont {Takesue}},\ }\bibfield  {title} {\bibinfo {title} {100,000-spin coherent {Ising} machine},\ }\href {https://doi.org/10.1126/sciadv.abh0952} {\bibfield  {journal} {\bibinfo  {journal} {Sci. Adv.}\ }\textbf {\bibinfo {volume} {7}},\ \bibinfo {pages}
  {eabh0952} (\bibinfo {year} {2021})}\BibitemShut {NoStop}%
\bibitem [{\citenamefont {Johnson}\ \emph {et~al.}(2011)\citenamefont {Johnson}, \citenamefont {Amin}, \citenamefont {Gildert}, \citenamefont {Lanting}, \citenamefont {Hamze}, \citenamefont {Dickson}, \citenamefont {Harris}, \citenamefont {Berkley}, \citenamefont {Johansson}, \citenamefont {Bunyk}, \citenamefont {Chapple}, \citenamefont {Enderud}, \citenamefont {Hilton}, \citenamefont {Karimi}, \citenamefont {Ladizinsky}, \citenamefont {Ladizinsky}, \citenamefont {Oh}, \citenamefont {Perminov}, \citenamefont {Rich}, \citenamefont {Thom}, \citenamefont {Tolkacheva}, \citenamefont {Truncik}, \citenamefont {Uchaikin}, \citenamefont {Wang}, \citenamefont {Wilson},\ and\ \citenamefont {Rose}}]{johnson_quantum_2011}%
  \BibitemOpen
  \bibfield  {author} {\bibinfo {author} {\bibfnamefont {M.~W.}\ \bibnamefont {Johnson}}, \bibinfo {author} {\bibfnamefont {M.~H.~S.}\ \bibnamefont {Amin}}, \bibinfo {author} {\bibfnamefont {S.}~\bibnamefont {Gildert}}, \bibinfo {author} {\bibfnamefont {T.}~\bibnamefont {Lanting}}, \bibinfo {author} {\bibfnamefont {F.}~\bibnamefont {Hamze}}, \bibinfo {author} {\bibfnamefont {N.}~\bibnamefont {Dickson}}, \bibinfo {author} {\bibfnamefont {R.}~\bibnamefont {Harris}}, \bibinfo {author} {\bibfnamefont {A.~J.}\ \bibnamefont {Berkley}}, \bibinfo {author} {\bibfnamefont {J.}~\bibnamefont {Johansson}}, \bibinfo {author} {\bibfnamefont {P.}~\bibnamefont {Bunyk}}, \bibinfo {author} {\bibfnamefont {E.~M.}\ \bibnamefont {Chapple}}, \bibinfo {author} {\bibfnamefont {C.}~\bibnamefont {Enderud}}, \bibinfo {author} {\bibfnamefont {J.~P.}\ \bibnamefont {Hilton}}, \bibinfo {author} {\bibfnamefont {K.}~\bibnamefont {Karimi}}, \bibinfo {author} {\bibfnamefont {E.}~\bibnamefont {Ladizinsky}}, \bibinfo {author} {\bibfnamefont
  {N.}~\bibnamefont {Ladizinsky}}, \bibinfo {author} {\bibfnamefont {T.}~\bibnamefont {Oh}}, \bibinfo {author} {\bibfnamefont {I.}~\bibnamefont {Perminov}}, \bibinfo {author} {\bibfnamefont {C.}~\bibnamefont {Rich}}, \bibinfo {author} {\bibfnamefont {M.~C.}\ \bibnamefont {Thom}}, \bibinfo {author} {\bibfnamefont {E.}~\bibnamefont {Tolkacheva}}, \bibinfo {author} {\bibfnamefont {C.~J.~S.}\ \bibnamefont {Truncik}}, \bibinfo {author} {\bibfnamefont {S.}~\bibnamefont {Uchaikin}}, \bibinfo {author} {\bibfnamefont {J.}~\bibnamefont {Wang}}, \bibinfo {author} {\bibfnamefont {B.}~\bibnamefont {Wilson}},\ and\ \bibinfo {author} {\bibfnamefont {G.}~\bibnamefont {Rose}},\ }\bibfield  {title} {\bibinfo {title} {Quantum annealing with manufactured spins},\ }\href {https://doi.org/10.1038/nature10012} {\bibfield  {journal} {\bibinfo  {journal} {Nature}\ }\textbf {\bibinfo {volume} {473}},\ \bibinfo {pages} {194} (\bibinfo {year} {2011})}\BibitemShut {NoStop}%
\bibitem [{\citenamefont {Pierangeli}\ \emph {et~al.}(2019)\citenamefont {Pierangeli}, \citenamefont {Marcucci},\ and\ \citenamefont {Conti}}]{pierangeli_large-scale_2019}%
  \BibitemOpen
  \bibfield  {author} {\bibinfo {author} {\bibfnamefont {D.}~\bibnamefont {Pierangeli}}, \bibinfo {author} {\bibfnamefont {G.}~\bibnamefont {Marcucci}},\ and\ \bibinfo {author} {\bibfnamefont {C.}~\bibnamefont {Conti}},\ }\bibfield  {title} {\bibinfo {title} {Large-scale photonic {Ising} machine by spatial light modulation},\ }\href {https://doi.org/10.1103/PhysRevLett.122.213902} {\bibfield  {journal} {\bibinfo  {journal} {Phys. Rev. Lett.}\ }\textbf {\bibinfo {volume} {122}},\ \bibinfo {pages} {213902} (\bibinfo {year} {2019})}\BibitemShut {NoStop}%
\bibitem [{\citenamefont {Yamashita}\ \emph {et~al.}(2023)\citenamefont {Yamashita}, \citenamefont {Okubo}, \citenamefont {Shimomura}, \citenamefont {Ogura}, \citenamefont {Tanida},\ and\ \citenamefont {Suzuki}}]{yamashita_low-rank_2023}%
  \BibitemOpen
  \bibfield  {author} {\bibinfo {author} {\bibfnamefont {H.}~\bibnamefont {Yamashita}}, \bibinfo {author} {\bibfnamefont {K.-i.}\ \bibnamefont {Okubo}}, \bibinfo {author} {\bibfnamefont {S.}~\bibnamefont {Shimomura}}, \bibinfo {author} {\bibfnamefont {Y.}~\bibnamefont {Ogura}}, \bibinfo {author} {\bibfnamefont {J.}~\bibnamefont {Tanida}},\ and\ \bibinfo {author} {\bibfnamefont {H.}~\bibnamefont {Suzuki}},\ }\bibfield  {title} {\bibinfo {title} {Low-rank combinatorial optimization and statistical learning by spatial photonic {Ising} machine},\ }\href {https://doi.org/10.1103/PhysRevLett.131.063801} {\bibfield  {journal} {\bibinfo  {journal} {Phys. Rev. Lett.}\ }\textbf {\bibinfo {volume} {131}},\ \bibinfo {pages} {063801} (\bibinfo {year} {2023})}\BibitemShut {NoStop}%
\bibitem [{\citenamefont {Fan}\ \emph {et~al.}(2023)\citenamefont {Fan}, \citenamefont {Sun}, \citenamefont {Xu}, \citenamefont {Wang}, \citenamefont {Zhu},\ and\ \citenamefont {Lin}}]{fan_programmable_2023}%
  \BibitemOpen
  \bibfield  {author} {\bibinfo {author} {\bibfnamefont {Y.}~\bibnamefont {Sun}}, \bibinfo {author} {\bibfnamefont {W.}~\bibnamefont {Fan}}, \bibinfo {author} {\bibfnamefont {X.}~\bibnamefont {Xu}}, \bibinfo {author} {\bibfnamefont {D.-W.}\ \bibnamefont {Wang}}, \bibinfo {author} {\bibfnamefont {S.-Y.}\ \bibnamefont {Zhu}},\ and\ \bibinfo {author} {\bibfnamefont {H.-Q.}\ \bibnamefont {Lin}},\ }\bibfield  {title} {\bibinfo {title} {Programmable photonic simulator for spin glass models},\ }\href {https://doi.org/10.1002/lpor.202402160} {\bibfield  {journal} {\bibinfo  {journal} {Laser \& Photonics Rev.}\ }\bibinfo {pages} {e02160} (\bibinfo {year} {2025})}\BibitemShut {NoStop}
\bibitem [{\citenamefont {Fang}\ \emph {et~al.}(2021)\citenamefont {Fang}, \citenamefont {Huang},\ and\ \citenamefont {Ruan}}]{fang_experimental_2021}%
  \BibitemOpen
  \bibfield  {author} {\bibinfo {author} {\bibfnamefont {Y.}~\bibnamefont {Fang}}, \bibinfo {author} {\bibfnamefont {J.}~\bibnamefont {Huang}},\ and\ \bibinfo {author} {\bibfnamefont {Z.}~\bibnamefont {Ruan}},\ }\bibfield  {title} {\bibinfo {title} {Experimental observation of phase transitions in spatial photonic {Ising} machine},\ }\href {https://doi.org/10.1103/PhysRevLett.127.043902} {\bibfield  {journal} {\bibinfo  {journal} {Phys. Rev. Lett.}\ }\textbf {\bibinfo {volume} {127}},\ \bibinfo {pages} {043902} (\bibinfo {year} {2021})}\BibitemShut {NoStop}%
\bibitem [{\citenamefont {Leonetti}\ \emph {et~al.}(2021)\citenamefont {Leonetti}, \citenamefont {Hörmann}, \citenamefont {Leuzzi}, \citenamefont {Parisi},\ and\ \citenamefont {Ruocco}}]{leonetti_optical_2021}%
  \BibitemOpen
  \bibfield  {author} {\bibinfo {author} {\bibfnamefont {M.}~\bibnamefont {Leonetti}}, \bibinfo {author} {\bibfnamefont {E.}~\bibnamefont {Hörmann}}, \bibinfo {author} {\bibfnamefont {L.}~\bibnamefont {Leuzzi}}, \bibinfo {author} {\bibfnamefont {G.}~\bibnamefont {Parisi}},\ and\ \bibinfo {author} {\bibfnamefont {G.}~\bibnamefont {Ruocco}},\ }\bibfield  {title} {\bibinfo {title} {Optical computation of a spin glass dynamics with tunable complexity},\ }\href {https://doi.org/10.1073/pnas.2015207118} {\bibfield  {journal} {\bibinfo  {journal} {Proc. Natl. Acad. Sci.}\ }\textbf {\bibinfo {volume} {118}},\ \bibinfo {pages} {e2015207118} (\bibinfo {year} {2021})}\BibitemShut {NoStop}%
\bibitem [{\citenamefont {Luo}\ \emph {et~al.}(2023)\citenamefont {Luo}, \citenamefont {Mi}, \citenamefont {Huang},\ and\ \citenamefont {Ruan}}]{luo_wavelength-division_2023}%
  \BibitemOpen
  \bibfield  {author} {\bibinfo {author} {\bibfnamefont {L.}~\bibnamefont {Luo}}, \bibinfo {author} {\bibfnamefont {Z.}~\bibnamefont {Mi}}, \bibinfo {author} {\bibfnamefont {J.}~\bibnamefont {Huang}},\ and\ \bibinfo {author} {\bibfnamefont {Z.}~\bibnamefont {Ruan}},\ }\bibfield  {title} {\bibinfo {title} {Wavelength-division multiplexing optical {Ising} simulator enabling fully programmable spin couplings and external magnetic fields},\ }\href {https://doi.org/10.1126/sciadv.adg6238} {\bibfield  {journal} {\bibinfo  {journal} {Science Advances}\ }\textbf {\bibinfo {volume} {9}},\ \bibinfo {pages} {eadg6238} (\bibinfo {year} {2023})}\BibitemShut {NoStop}%
\bibitem [{\citenamefont {Ouyang}\ \emph {et~al.}(2024)\citenamefont {Ouyang}, \citenamefont {Liao}, \citenamefont {Feng}, \citenamefont {Li}, \citenamefont {Cui}, \citenamefont {Liu}, \citenamefont {Sun}, \citenamefont {Zhang},\ and\ \citenamefont {Huang}}]{ouyang_programmable_2024}%
  \BibitemOpen
  \bibfield  {author} {\bibinfo {author} {\bibfnamefont {J.}~\bibnamefont {Ouyang}}, \bibinfo {author} {\bibfnamefont {Y.}~\bibnamefont {Liao}}, \bibinfo {author} {\bibfnamefont {X.}~\bibnamefont {Feng}}, \bibinfo {author} {\bibfnamefont {Y.}~\bibnamefont {Li}}, \bibinfo {author} {\bibfnamefont {K.}~\bibnamefont {Cui}}, \bibinfo {author} {\bibfnamefont {F.}~\bibnamefont {Liu}}, \bibinfo {author} {\bibfnamefont {H.}~\bibnamefont {Sun}}, \bibinfo {author} {\bibfnamefont {W.}~\bibnamefont {Zhang}},\ and\ \bibinfo {author} {\bibfnamefont {Y.}~\bibnamefont {Huang}},\ }\bibfield  {title} {\bibinfo {title} {Programmable and reconfigurable photonic simulator for classical {XY} models},\ }\href {https://doi.org/10.1103/PhysRevApplied.22.L021001} {\bibfield  {journal} {\bibinfo  {journal} {Phys. Rev. Appl.} \ }\textbf {\bibinfo {volume} {22}},\ \bibinfo {pages} {L021001} (\bibinfo {year} {2024})}\BibitemShut {NoStop}
\bibitem [{\citenamefont {Nixon}\ \emph {et~al.}(2013)\citenamefont {Nixon}, \citenamefont {Ronen}, \citenamefont {Friesem},\ and\ \citenamefont {Davidson}}]{nixon_observing_2013}%
  \BibitemOpen
  \bibfield  {author} {\bibinfo {author} {\bibfnamefont {M.}~\bibnamefont {Nixon}}, \bibinfo {author} {\bibfnamefont {E.}~\bibnamefont {Ronen}}, \bibinfo {author} {\bibfnamefont {A.~A.}\ \bibnamefont {Friesem}},\ and\ \bibinfo {author} {\bibfnamefont {N.}~\bibnamefont {Davidson}},\ }\bibfield  {title} {\bibinfo {title} {Observing geometric frustration with thousands of coupled lasers},\ }\href {https://doi.org/10.1103/PhysRevLett.110.184102} {\bibfield  {journal} {\bibinfo  {journal} {Phys. Rev. Lett.}\ }\textbf {\bibinfo {volume} {110}},\ \bibinfo {pages} {184102} (\bibinfo {year} {2013})}\BibitemShut {NoStop}%
\bibitem [{\citenamefont {Gershenzon}\ \emph {et~al.}(2020)\citenamefont {Gershenzon}, \citenamefont {Arwas}, \citenamefont {Gadasi}, \citenamefont {Tradonsky}, \citenamefont {Friesem}, \citenamefont {Raz},\ and\ \citenamefont {Davidson}}]{gershenzon_exact_2020}%
  \BibitemOpen
  \bibfield  {author} {\bibinfo {author} {\bibfnamefont {I.}~\bibnamefont {Gershenzon}}, \bibinfo {author} {\bibfnamefont {G.}~\bibnamefont {Arwas}}, \bibinfo {author} {\bibfnamefont {S.}~\bibnamefont {Gadasi}}, \bibinfo {author} {\bibfnamefont {C.}~\bibnamefont {Tradonsky}}, \bibinfo {author} {\bibfnamefont {A.}~\bibnamefont {Friesem}}, \bibinfo {author} {\bibfnamefont {O.}~\bibnamefont {Raz}},\ and\ \bibinfo {author} {\bibfnamefont {N.}~\bibnamefont {Davidson}},\ }\bibfield  {title} {\bibinfo {title} {Exact mapping between a laser network loss rate and the classical {XY} {Hamiltonian} by laser loss control},\ }\href {https://doi.org/10.1515/nanoph-2020-0137} {\bibfield  {journal} {\bibinfo  {journal} {Nanophotonics}\ }\textbf {\bibinfo {volume} {9}},\ \bibinfo {pages} {4117} (\bibinfo {year} {2020})}\BibitemShut {NoStop}%
\bibitem [{\citenamefont {Parto}\ \emph {et~al.}(2020)\citenamefont {Parto}, \citenamefont {Hayenga}, \citenamefont {Marandi},\ \emph {et~al.}}]{parto_realizing_2020}%
  \BibitemOpen
  \bibfield  {author} {\bibinfo {author} {\bibfnamefont {M.}~\bibnamefont {Parto}}, \bibinfo {author} {\bibfnamefont {W.}~\bibnamefont {Hayenga}}, \bibinfo {author} {\bibfnamefont {A.}~\bibnamefont {Marandi}}, \bibinfo {author} {\bibfnamefont {D.~N.}~\bibnamefont {Christodoulides}},\ and\ \bibinfo {author} {\bibfnamefont {M.}~\bibnamefont {Khajavikhan}},\ }\bibfield  {title} {\bibinfo {title} {Realizing spin Hamiltonians in nanoscale active photonic lattices},\ }\href {https://doi.org/10.1038/s41563-020-0635-6} {\bibfield  {journal} {\bibinfo  {journal} {Nat. Mater.}\ }\textbf {\bibinfo {volume} {19}},\ \bibinfo {pages} {725} (\bibinfo {year} {2020})}\BibitemShut {NoStop}%
\bibitem [{\citenamefont {Berloff}\ \emph {et~al.}(2017)\citenamefont {Berloff}, \citenamefont {Silva}, \citenamefont {Kalinin}, \citenamefont {Askitopoulos}, \citenamefont {Töpfer}, \citenamefont {Cilibrizzi}, \citenamefont {Langbein},\ and\ \citenamefont {Lagoudakis}}]{berloff_realizing_2017}%
  \BibitemOpen
  \bibfield  {author} {\bibinfo {author} {\bibfnamefont {N.~G.}\ \bibnamefont {Berloff}}, \bibinfo {author} {\bibfnamefont {M.}~\bibnamefont {Silva}}, \bibinfo {author} {\bibfnamefont {K.}~\bibnamefont {Kalinin}}, \bibinfo {author} {\bibfnamefont {A.}~\bibnamefont {Askitopoulos}}, \bibinfo {author} {\bibfnamefont {J.~D.}\ \bibnamefont {Töpfer}}, \bibinfo {author} {\bibfnamefont {P.}~\bibnamefont {Cilibrizzi}}, \bibinfo {author} {\bibfnamefont {W.}~\bibnamefont {Langbein}},\ and\ \bibinfo {author} {\bibfnamefont {P.~G.}\ \bibnamefont {Lagoudakis}},\ }\bibfield  {title} {\bibinfo {title} {Realizing the classical {XY} {Hamiltonian} in polariton simulators},\ }\href {https://doi.org/10.1038/nmat4971} {\bibfield  {journal} {\bibinfo  {journal} {Nat. Mater.}\ }\textbf {\bibinfo {volume} {16}},\ \bibinfo {pages} {1120} (\bibinfo {year} {2017})}\BibitemShut {NoStop}%
\bibitem [{\citenamefont {Kalinin}\ \emph {et~al.}(2020)\citenamefont {Kalinin}, \citenamefont {Amo}, \citenamefont {Bloch},\ and\ \citenamefont {Berloff}}]{kalinin_polaritonic_2020}%
  \BibitemOpen
  \bibfield  {author} {\bibinfo {author} {\bibfnamefont {K.~P.}\ \bibnamefont {Kalinin}}, \bibinfo {author} {\bibfnamefont {A.}~\bibnamefont {Amo}}, \bibinfo {author} {\bibfnamefont {J.}~\bibnamefont {Bloch}},\ and\ \bibinfo {author} {\bibfnamefont {N.~G.}\ \bibnamefont {Berloff}},\ }\bibfield  {title} {\bibinfo {title} {Polaritonic {XY}-{Ising} machine},\ }\href {https://doi.org/10.1515/nanoph-2020-0162} {\bibfield  {journal} {\bibinfo  {journal} {Nanophotonics}\ }\textbf {\bibinfo {volume} {9}},\ \bibinfo {pages} {4127} (\bibinfo {year} {2020})}\BibitemShut {NoStop}%
\bibitem [{\citenamefont {Feng}\ \emph {et~al.}(2024)\citenamefont {Feng}, \citenamefont {Li}, \citenamefont {Yuan}, \citenamefont {Hasman}, \citenamefont {Wang},\ and\ \citenamefont {Chen}}]{feng_Spin_2024}%
  \BibitemOpen
  \bibfield  {author} {\bibinfo {author} {\bibfnamefont {J.}~\bibnamefont {Feng}}, \bibinfo {author} {\bibfnamefont {Z.}~\bibnamefont {Li}}, \bibinfo {author} {\bibfnamefont {L.}~\bibnamefont {Yuan}}, \bibinfo {author} {\bibfnamefont {E.}~\bibnamefont {Hasman}}, \bibinfo {author} {\bibfnamefont {B.}~\bibnamefont {Wang}},\ and\ \bibinfo {author} {\bibfnamefont {X.}~\bibnamefont {Chen}},\ }\bibfield  {title} {\bibinfo {title} {{Spin Hamiltonians in the modulated momenta of light}},\ }\href {https://doi.org/10.1117/1.AP.7.4.046001} {\bibfield  {journal} {\bibinfo  {journal} {Adv. Photonics}\ }\textbf {\bibinfo {volume} {7}},\ \bibinfo {pages} {046001} (\bibinfo {year} {2025})}\BibitemShut {NoStop}
\bibitem [{\citenamefont {Yu}\ \emph {et~al.}(2024)\citenamefont {Yu}, \citenamefont {He}, \citenamefont {Fang}, \citenamefont {Deng},\ and\ \citenamefont {Yuan}}]{yuSpatialOpticalSimulator2024}%
  \BibitemOpen
  \bibfield  {author} {\bibinfo {author} {\bibfnamefont {S.-T.}\ \bibnamefont {Yu}}, \bibinfo {author} {\bibfnamefont {M.-G.}\ \bibnamefont {He}}, \bibinfo {author} {\bibfnamefont {S.}~\bibnamefont {Fang}}, \bibinfo {author} {\bibfnamefont {Y.}~\bibnamefont {Deng}},\ and\ \bibinfo {author} {\bibfnamefont {Z.-S.}\ \bibnamefont {Yuan}},\ }\bibfield  {title} {\bibinfo {title} {Spatial optical simulator for classical statistical models},\ }\href {https://doi.org/10.1103/PhysRevLett.133.237101} {\bibfield  {journal} {\bibinfo  {journal} {Phys. Rev. Lett.}\ }\textbf {\bibinfo {volume} {133}},\ \bibinfo {pages} {237101} (\bibinfo {year} {2024})}\BibitemShut {NoStop}%
\bibitem [{\citenamefont {Cavaliere}\ \emph {et~al.}(2021)\citenamefont {Cavaliere}, \citenamefont {Lesieur},\ and\ \citenamefont {Ricci-Tersenghi}}]{cavaliere_optimization_2021}%
  \BibitemOpen
  \bibfield  {author} {\bibinfo {author} {\bibfnamefont {A.~G.}~\bibnamefont {Cavaliere}}, \bibinfo {author} {\bibfnamefont {T.}~\bibnamefont {Lesieur}},\ and\ \bibinfo {author} {\bibfnamefont {F.}~\bibnamefont {Ricci-Tersenghi}},\ }\bibfield  {title} {\bibinfo {title} {Optimization of the dynamic transition in the continuous coloring problem},\ }\href {https://doi.org/10.1088/1742-5468/ac382e} {\bibfield  {journal} {\bibinfo  {journal} {J. Stat. Mech.}\ }\textbf {\bibinfo {volume} {2021}}, \bibinfo {pages} {113302} (\bibinfo {year} {2021})}\BibitemShut {NoStop}%
\bibitem [{\citenamefont {Singer}(2011)\citenamefont {Singer}}]{singer_angular_2011}%
  \BibitemOpen
  \bibfield  {author} {\bibinfo {author} {\bibfnamefont {A.}~\bibnamefont {Singer}},\ }\bibfield  {title} {\bibinfo {title} {Angular synchronization by eigenvectors and semidefinite programming},\ }\href {https://doi.org/10.1016/j.acha.2010.02.001} {\bibfield  {journal} {\bibinfo  {journal} {Appl. Comput. Harmon. Anal.}\ }\textbf {\bibinfo {volume} {30}}, \bibinfo {pages} {20} (\bibinfo {year} {2011})}\BibitemShut {NoStop}%
\bibitem [{\citenamefont {Harrison}\ \emph {et~al.}(2022)\citenamefont {Harrison}, \citenamefont {Sigurdsson}, \citenamefont {Alyatkin}, \citenamefont {T\"opfer},\ and\ \citenamefont {Lagoudakis}}]{PhysRevApplied.17.024063}%
  \BibitemOpen
  \bibfield  {author} {\bibinfo {author} {\bibfnamefont {S.~L.}~\bibnamefont {Harrison}}, \bibinfo {author} {\bibfnamefont {H.}~\bibnamefont {Sigurdsson}}, \bibinfo {author} {\bibfnamefont {S.}~\bibnamefont {Alyatkin}}, \bibinfo {author} {\bibfnamefont {J.~D.}~\bibnamefont {T\"opfer}},\ and\ \bibinfo {author} {\bibfnamefont {P.~G.}~\bibnamefont {Lagoudakis}},\ }\bibfield  {title} {\bibinfo {title} {Solving the Max-3-Cut Problem with Coherent Networks},\ }\href {https://doi.org/10.1103/PhysRevApplied.17.024063} {\bibfield  {journal} {\bibinfo  {journal} {Phys. Rev. Appl.}\ }\textbf {\bibinfo {volume} {17}}, \bibinfo {pages} {024063} (\bibinfo {year} {2022})}\BibitemShut {NoStop}%
\bibitem [{\citenamefont {Shaginyan}\ \emph {et~al.}(2020)\citenamefont {Shaginyan}, \citenamefont {Stephanovich}, \citenamefont {Msezane}, \citenamefont {Japaridze}, \citenamefont {Clark}, \citenamefont {Amusia},\ and\ \citenamefont {Kirichenko}}]{shaginyan_theoretical_2020}%
  \BibitemOpen
  \bibfield  {author} {\bibinfo {author} {\bibfnamefont {V.~R.}\ \bibnamefont {Shaginyan}}, \bibinfo {author} {\bibfnamefont {V.~A.}\ \bibnamefont {Stephanovich}}, \bibinfo {author} {\bibfnamefont {A.~Z.}\ \bibnamefont {Msezane}}, \bibinfo {author} {\bibfnamefont {G.~S.}\ \bibnamefont {Japaridze}}, \bibinfo {author} {\bibfnamefont {J.~W.}\ \bibnamefont {Clark}}, \bibinfo {author} {\bibfnamefont {M.~Y.}\ \bibnamefont {Amusia}},\ and\ \bibinfo {author} {\bibfnamefont {E.~V.}\ \bibnamefont {Kirichenko}},\ }\bibfield  {title} {\bibinfo {title} {Theoretical and experimental developments in quantum spin liquid in geometrically frustrated magnets: {A} review},\ }\href {https://doi.org/10.1007/s10853-019-04128-w} {\bibfield  {journal} {\bibinfo  {journal} {J. Mater. Sci.}\ }\textbf {\bibinfo {volume} {55}},\ \bibinfo {pages} {2257} (\bibinfo {year} {2020})}\BibitemShut {NoStop}%
\bibitem [{\citenamefont {Kawamura}(2011)}]{kawamura_ordering_2011}%
  \BibitemOpen
  \bibfield  {author} {\bibinfo {author} {\bibfnamefont {H.}~\bibnamefont {Kawamura}},\ }\bibfield  {title} {\bibinfo {title} {The ordering of \textit{{XY}} spin glasses},\ }\href {https://doi.org/10.1088/0953-8984/23/16/164210} {\bibfield  {journal} {\bibinfo  {journal} {J. Phys.: Condens. Matter}\ }\textbf {\bibinfo {volume} {23}},\ \bibinfo {pages} {164210} (\bibinfo {year} {2011})}\BibitemShut {NoStop}%
\bibitem [{\citenamefont {Mattis}(1976)}]{mattis_solvable_1976}%
  \BibitemOpen
  \bibfield  {author} {\bibinfo {author} {\bibfnamefont {D.~C.}\ \bibnamefont {Mattis}},\ }\bibfield  {title} {\bibinfo {title} {Solvable spin systems with random interactions},\ }\href {https://doi.org/10.1016/0375-9601(76)90396-0} {\bibfield  {journal} {\bibinfo  {journal} {Phys. Lett. A}\ }\textbf {\bibinfo {volume} {56}},\ \bibinfo {pages} {421} (\bibinfo {year} {1976})}\BibitemShut {NoStop}%
\bibitem [{sup()}]{supp}%
  \BibitemOpen
  \href@noop {} {}\bibinfo {note} {See Supplementary Materials at \url{http://link.aps.org/supplemental/10.1103/tgt8-gb13} for detailed derivation, which includes Refs. \cite{fang_experimental_2021,goodman_introduction_2005,fan_programmable_2023,Goorden_superpixel-based_2014,jesacher_wavefront_2007,bergman_order-by-disorder_2007,di_ciolo_spiral_2014,muller_fast_2023,giachetti_berezinskii-kosterlitz-thouless_2021,xiao_two-dimensional_2024, kosterlitz_nobel_2017, kosterlitz_critical_1974}. All data used to produce the findings of this study are available via Zenodo at \url{https://doi.org/10.5281/zenodo.15120839}}\BibitemShut {Stop}%
\bibitem [{\citenamefont {Bergman}\ \emph {et~al.}(2007)\citenamefont {Bergman}, \citenamefont {Alicea}, \citenamefont {Gull}, \citenamefont {Trebst},\ and\ \citenamefont {Balents}}]{bergman_order-by-disorder_2007}%
  \BibitemOpen
  \bibfield  {author} {\bibinfo {author} {\bibfnamefont {D.}~\bibnamefont {Bergman}}, \bibinfo {author} {\bibfnamefont {J.}~\bibnamefont {Alicea}}, \bibinfo {author} {\bibfnamefont {E.}~\bibnamefont {Gull}}, \bibinfo {author} {\bibfnamefont {S.}~\bibnamefont {Trebst}},\ and\ \bibinfo {author} {\bibfnamefont {L.}~\bibnamefont {Balents}},\ }\bibfield  {title} {\bibinfo {title} {Order-by-disorder and spiral spin-liquid in frustrated diamond-lattice antiferromagnets},\ }\href {https://doi.org/10.1038/nphys622} {\bibfield  {journal} {\bibinfo  {journal} {Nat. Phys.}\ }\textbf {\bibinfo {volume} {3}},\ \bibinfo {pages} {487} (\bibinfo {year} {2007})}\BibitemShut {NoStop}%
\bibitem [{\citenamefont {Di~Ciolo}\ \emph {et~al.}(2014)\citenamefont {Di~Ciolo}, \citenamefont {Carrasquilla}, \citenamefont {Becca}, \citenamefont {Rigol},\ and\ \citenamefont {Galitski}}]{di_ciolo_spiral_2014}%
  \BibitemOpen
  \bibfield  {author} {\bibinfo {author} {\bibfnamefont {A.}~\bibnamefont {Di~Ciolo}}, \bibinfo {author} {\bibfnamefont {J.}~\bibnamefont {Carrasquilla}}, \bibinfo {author} {\bibfnamefont {F.}~\bibnamefont {Becca}}, \bibinfo {author} {\bibfnamefont {M.}~\bibnamefont {Rigol}},\ and\ \bibinfo {author} {\bibfnamefont {V.}~\bibnamefont {Galitski}},\ }\bibfield  {title} {\bibinfo {title} {Spiral antiferromagnets beyond the spin-wave approximation: {Frustrated} {XY} and {Heisenberg} models on the honeycomb lattice},\ }\href {https://doi.org/10.1103/PhysRevB.89.094413} {\bibfield  {journal} {\bibinfo  {journal} {Phys. Rev. B}\ }\textbf {\bibinfo {volume} {89}},\ \bibinfo {pages} {094413} (\bibinfo {year} {2014})}\BibitemShut {NoStop}%
\bibitem [{\citenamefont {Müller}\ \emph {et~al.}(2023)\citenamefont {Müller}, \citenamefont {Christiansen}, \citenamefont {Schnabel},\ and\ \citenamefont {Janke}}]{muller_fast_2023}%
  \BibitemOpen
  \bibfield  {author} {\bibinfo {author} {\bibfnamefont {F.}~\bibnamefont {Müller}}, \bibinfo {author} {\bibfnamefont {H.}~\bibnamefont {Christiansen}}, \bibinfo {author} {\bibfnamefont {S.}~\bibnamefont {Schnabel}},\ and\ \bibinfo {author} {\bibfnamefont {W.}~\bibnamefont {Janke}},\ }\bibfield  {title} {\bibinfo {title} {Fast, hierarchical, and adaptive algorithm for {Metropolis} {Monte} {Carlo} simulations of long-range interacting systems},\ }\href {https://doi.org/10.1103/PhysRevX.13.031006} {\bibfield  {journal} {\bibinfo  {journal} {Phys. Rev. X}\ }\textbf {\bibinfo {volume} {13}},\ \bibinfo {pages} {031006} (\bibinfo {year} {2023})}\BibitemShut {NoStop}%
\bibitem [{\citenamefont {Giachetti}\ \emph {et~al.}(2021)\citenamefont {Giachetti}, \citenamefont {Defenu}, \citenamefont {Ruffo},\ and\ \citenamefont {Trombettoni}}]{giachetti_berezinskii-kosterlitz-thouless_2021}%
  \BibitemOpen
  \bibfield  {author} {\bibinfo {author} {\bibfnamefont {G.}~\bibnamefont {Giachetti}}, \bibinfo {author} {\bibfnamefont {N.}~\bibnamefont {Defenu}}, \bibinfo {author} {\bibfnamefont {S.}~\bibnamefont {Ruffo}},\ and\ \bibinfo {author} {\bibfnamefont {A.}~\bibnamefont {Trombettoni}},\ }\bibfield  {title} {\bibinfo {title} {Berezinskii-{Kosterlitz}-{Thouless} phase transitions with long-range couplings},\ }\href {https://doi.org/10.1103/PhysRevLett.127.156801} {\bibfield  {journal} {\bibinfo  {journal} {Phys. Rev. Lett.}\ }\textbf {\bibinfo {volume} {127}},\ \bibinfo {pages} {156801} (\bibinfo {year} {2021})}\BibitemShut {NoStop}%
\bibitem [{\citenamefont {Goodman}(2005)}]{goodman_introduction_2005}%
  \BibitemOpen
  \bibfield  {author} {\bibinfo {author} {\bibfnamefont {J.~W.}\ \bibnamefont {Goodman}},\ }\href@noop {} {\emph {\bibinfo {title} {Introduction to {Fourier} {Optics}}}}\ (\bibinfo  {publisher} {Roberts and Company Publishers},\ \bibinfo {year} {2005})\BibitemShut {NoStop}%
\bibitem [{\citenamefont {Goorden}\ \emph {et~al.}(2014)\citenamefont {Goorden}, \citenamefont {Bertolotti},\ and\ \citenamefont {Mosk}}]{Goorden_superpixel-based_2014}%
  \BibitemOpen
  \bibfield  {author} {\bibinfo {author} {\bibfnamefont {S.~A.}\ \bibnamefont {Goorden}}, \bibinfo {author} {\bibfnamefont {J.}~\bibnamefont {Bertolotti}},\ and\ \bibinfo {author} {\bibfnamefont {A.~P.}\ \bibnamefont {Mosk}},\ }\bibfield  {title} {\bibinfo {title} {Superpixel-based spatial amplitude and phase modulation using a digital micromirror device},\ }\href {https://doi.org/10.1364/OE.22.017999} {\bibfield  {journal} {\bibinfo  {journal} {Opt. Express}\ }\textbf {\bibinfo {volume} {22}},\ \bibinfo {pages} {17999} (\bibinfo {year} {2014})}\BibitemShut {NoStop}%
\bibitem [{\citenamefont {Jesacher}\ \emph {et~al.}(2007)\citenamefont {Jesacher}, \citenamefont {Schwaighofer}, \citenamefont {F\"{u}rhapter}, \citenamefont {Maurer}, \citenamefont {Bernet},\ and\ \citenamefont {Ritsch-Marte}}]{jesacher_wavefront_2007}%
  \BibitemOpen
  \bibfield  {author} {\bibinfo {author} {\bibfnamefont {A.}~\bibnamefont {Jesacher}}, \bibinfo {author} {\bibfnamefont {A.}~\bibnamefont {Schwaighofer}}, \bibinfo {author} {\bibfnamefont {S.}~\bibnamefont {F\"{u}rhapter}}, \bibinfo {author} {\bibfnamefont {C.}~\bibnamefont {Maurer}}, \bibinfo {author} {\bibfnamefont {S.}~\bibnamefont {Bernet}},\ and\ \bibinfo {author} {\bibfnamefont {M.}~\bibnamefont {Ritsch-Marte}},\ }\bibfield  {title} {\bibinfo {title} {Wavefront correction of spatial light modulators using an optical vortex image},\ }\href {https://doi.org/10.1364/OE.15.005801} {\bibfield  {journal} {\bibinfo  {journal} {Opt. Express}\ }\textbf {\bibinfo {volume} {15}},\ \bibinfo {pages} {5801} (\bibinfo {year} {2007})}\BibitemShut {NoStop}%
\bibitem [{\citenamefont {Xiao}\ \emph {et~al.}(2024)\citenamefont {Xiao}, \citenamefont {Yao}, \citenamefont {Zhang}, \citenamefont {Fan},\ and\ \citenamefont {Deng}}]{xiao_two-dimensional_2024}%
  \BibitemOpen
  \bibfield  {author} {\bibinfo {author} {\bibfnamefont {T.}~\bibnamefont {Xiao}}, \bibinfo {author} {\bibfnamefont {D.}~\bibnamefont {Yao}}, \bibinfo {author} {\bibfnamefont {C.}~\bibnamefont {Zhang}}, \bibinfo {author} {\bibfnamefont {Z.}~\bibnamefont {Fan}},\ and\ \bibinfo {author} {\bibfnamefont {Y.}~\bibnamefont {Deng}},\ }\bibfield  {title} {\bibinfo {title} {Two-dimensional {XY} ferromagnet induced by long-range interaction},\ }\href {https://doi.org/10.1088/0256-307X/42/7/070002} {\bibfield  {journal} {\bibinfo  {journal} {Chin. Phys. Lett.}\ }\textbf {\bibinfo {volume} {42}},\ \bibinfo {pages} {070002} (\bibinfo {year} {2025})}\BibitemShut {NoStop}
\bibitem [{\citenamefont {Kosterlitz}(2017)}]{kosterlitz_nobel_2017}%
  \BibitemOpen
  \bibfield  {author} {\bibinfo {author} {\bibfnamefont {J.~M.}\ \bibnamefont {Kosterlitz}},\ }\bibfield  {title} {\bibinfo {title} {Nobel {Lecture}: {Topological} defects and phase transitions},\ }\href {https://doi.org/10.1103/RevModPhys.89.040501} {\bibfield  {journal} {\bibinfo  {journal} {Rev. Mod. Phys.}\ }\textbf {\bibinfo {volume} {89}},\ \bibinfo {pages} {040501} (\bibinfo {year} {2017})}\BibitemShut {NoStop}%
\bibitem [{\citenamefont {Chib}\ and\ \citenamefont {Greenberg}(1995)}]{chib_understanding_1995}%
  \BibitemOpen
  \bibfield  {author} {\bibinfo {author} {\bibfnamefont {S.}~\bibnamefont {Chib}}\ and\ \bibinfo {author} {\bibfnamefont {E.}~\bibnamefont {Greenberg}},\ }\bibfield  {title} {\bibinfo {title} {Understanding the {Metropolis}-{Hastings} algorithm},\ }\href {https://doi.org/10.1080/00031305.1995.10476177} {\bibfield  {journal} {\bibinfo  {journal} {Am. Stat.}\ }\textbf {\bibinfo {volume} {49}},\ \bibinfo {pages} {327} (\bibinfo {year} {1995})}\BibitemShut {NoStop}%
\bibitem [{\citenamefont {Earl}\ and\ \citenamefont {Deem}(2005)}]{earl_parallel_2005}%
  \BibitemOpen
  \bibfield  {author} {\bibinfo {author} {\bibfnamefont {D.~J.}\ \bibnamefont {Earl}}\ and\ \bibinfo {author} {\bibfnamefont {M.~W.}\ \bibnamefont {Deem}},\ }\bibfield  {title} {\bibinfo {title} {Parallel tempering: {Theory}, applications, and new perspectives},\ }\href {https://doi.org/10.1039/B509983H} {\bibfield  {journal} {\bibinfo  {journal} {Phys. Chem. Chem. Phys.}\ }\textbf {\bibinfo {volume} {7}},\ \bibinfo {pages} {3910} (\bibinfo {year} {2005})}\BibitemShut {NoStop}%
\bibitem [{\citenamefont {Mermin}\ and\ \citenamefont {Wagner}(1966)}]{mermin_absence_1966}%
  \BibitemOpen
  \bibfield  {author} {\bibinfo {author} {\bibfnamefont {N.~D.}\ \bibnamefont {Mermin}}\ and\ \bibinfo {author} {\bibfnamefont {H.}~\bibnamefont {Wagner}},\ }\bibfield  {title} {\bibinfo {title} {Absence of ferromagnetism or antiferromagnetism in one- or two-dimensional isotropic {Heisenberg} models},\ }\href {https://doi.org/10.1103/PhysRevLett.17.1133} {\bibfield  {journal} {\bibinfo  {journal} {Phys. Rev. Lett.}\ }\textbf {\bibinfo {volume} {17}},\ \bibinfo {pages} {1133} (\bibinfo {year} {1966})}\BibitemShut {NoStop}%
\bibitem [{\citenamefont {Hohenberg}(1967)}]{hohenberg_existence_1967}%
  \BibitemOpen
  \bibfield  {author} {\bibinfo {author} {\bibfnamefont {P.~C.}\ \bibnamefont {Hohenberg}},\ }\bibfield  {title} {\bibinfo {title} {Existence of long-range order in one and two dimensions},\ }\href {https://doi.org/10.1103/PhysRev.158.383} {\bibfield  {journal} {\bibinfo  {journal} {Phys. Rev.}\ }\textbf {\bibinfo {volume} {158}},\ \bibinfo {pages} {383} (\bibinfo {year} {1967})}\BibitemShut {NoStop}%
\bibitem [{\citenamefont {Zinn-Justin}(2021)}]{zinn-justin_quantum_2021}%
  \BibitemOpen
  \bibfield  {author} {\bibinfo {author} {\bibfnamefont {J.}~\bibnamefont {Zinn-Justin}},\ }\href@noop {} {\emph {\bibinfo {title} {Quantum {Field} {Theory} and {Critical} {Phenomena}: {Fifth} {Edition}}}}\ (\bibinfo  {publisher} {Oxford University Press},\ \bibinfo {year} {2021})\BibitemShut {NoStop}%
\bibitem [{\citenamefont {Kosterlitz}\ and\ \citenamefont {Thouless}(1972)}]{kosterlitz_long_1972}%
  \BibitemOpen
  \bibfield  {author} {\bibinfo {author} {\bibfnamefont {J.~M.}\ \bibnamefont {Kosterlitz}}\ and\ \bibinfo {author} {\bibfnamefont {D.~J.}\ \bibnamefont {Thouless}},\ }\bibfield  {title} {\bibinfo {title} {Long range order and metastability in two dimensional solids and superfluids. ({Application} of dislocation theory)},\ }\href {https://doi.org/10.1088/0022-3719/5/11/002} {\bibfield  {journal} {\bibinfo  {journal} {J. Phys. C: Solid State Phys.}\ }\textbf {\bibinfo {volume} {5}},\ \bibinfo {pages} {L124} (\bibinfo {year} {1972})}\BibitemShut {NoStop}%
\bibitem [{\citenamefont {Weber}\ and\ \citenamefont {Minnhagen}(1988)}]{weber_monte_1988}%
  \BibitemOpen
  \bibfield  {author} {\bibinfo {author} {\bibfnamefont {H.}~\bibnamefont {Weber}}\ and\ \bibinfo {author} {\bibfnamefont {P.}~\bibnamefont {Minnhagen}},\ }\bibfield  {title} {\bibinfo {title} {Monte {Carlo} determination of the critical temperature for the two-dimensional {XY} model},\ }\href {https://doi.org/10.1103/PhysRevB.37.5986} {\bibfield  {journal} {\bibinfo  {journal} {Phys. Rev. B}\ }\textbf {\bibinfo {volume} {37}},\ \bibinfo {pages} {5986} (\bibinfo {year} {1988})}\BibitemShut {NoStop}%
\bibitem [{\citenamefont {Situ}\ and\ \citenamefont {Fleischer}(2020)}]{situ_dynamics_2020}%
  \BibitemOpen
  \bibfield  {author} {\bibinfo {author} {\bibfnamefont {G.}~\bibnamefont {Situ}}\ and\ \bibinfo {author} {\bibfnamefont {J.~W.}\ \bibnamefont {Fleischer}},\ }\bibfield  {title} {\bibinfo {title} {Dynamics of the {Berezinskii}–{Kosterlitz}–{Thouless} transition in a photon fluid},\ }\href {https://doi.org/10.1038/s41566-020-0636-7} {\bibfield  {journal} {\bibinfo  {journal} {Nat. Photon.}\ }\textbf {\bibinfo {volume} {14}},\ \bibinfo {pages} {517} (\bibinfo {year} {2020})}\BibitemShut {NoStop}%
\bibitem [{\citenamefont {Lee}\ and\ \citenamefont {Lee}(1998)}]{lee_phase_1998}%
  \BibitemOpen
  \bibfield  {author} {\bibinfo {author} {\bibfnamefont {S.}~\bibnamefont {Lee}}\ and\ \bibinfo {author} {\bibfnamefont {K.-C.}\ \bibnamefont {Lee}},\ }\bibfield  {title} {\bibinfo {title} {Phase transitions in the fully frustrated triangular {XY} model},\ }\href {https://doi.org/10.1103/PhysRevB.57.8472} {\bibfield  {journal} {\bibinfo  {journal} {Phys. Rev. B}\ }\textbf {\bibinfo {volume} {57}},\ \bibinfo {pages} {8472} (\bibinfo {year} {1998})}\BibitemShut {NoStop}%
\bibitem [{\citenamefont {Olsson}(1995)}]{olsson_two_1995}%
  \BibitemOpen
  \bibfield  {author} {\bibinfo {author} {\bibfnamefont {P.}~\bibnamefont {Olsson}},\ }\bibfield  {title} {\bibinfo {title} {Two phase transitions in the fully frustrated $\mathit{XY}$ model},\ }\href {https://doi.org/10.1103/PhysRevLett.75.2758} {\bibfield  {journal} {\bibinfo  {journal} {Phys. Rev. Lett.}\ }\textbf {\bibinfo {volume} {75}},\ \bibinfo {pages} {2758} (\bibinfo {year} {1995})}\BibitemShut {NoStop}%
\bibitem [{\citenamefont {Minnhagen}(1985)}]{minnhagen_nonuniversal_1985}%
  \BibitemOpen
  \bibfield  {author} {\bibinfo {author} {\bibfnamefont {P.}~\bibnamefont {Minnhagen}},\ }\bibfield  {title} {\bibinfo {title} {Nonuniversal jumps and the {Kosterlitz}-{Thouless} transition},\ }\href {https://doi.org/10.1103/PhysRevLett.54.2351} {\bibfield  {journal} {\bibinfo  {journal} {Phys. Rev. Lett.}\ }\textbf {\bibinfo {volume} {54}},\ \bibinfo {pages} {2351} (\bibinfo {year} {1985})}\BibitemShut {NoStop}%
\bibitem [{\citenamefont {Yosefin}\ and\ \citenamefont {Domany}(1985)}]{yosefin_phase_1985}%
  \BibitemOpen
  \bibfield  {author} {\bibinfo {author} {\bibfnamefont {M.}~\bibnamefont {Yosefin}}\ and\ \bibinfo {author} {\bibfnamefont {E.}~\bibnamefont {Domany}},\ }\bibfield  {title} {\bibinfo {title} {Phase transitions in fully frustrated spin systems},\ }\href {https://doi.org/10.1103/PhysRevB.32.1778} {\bibfield  {journal} {\bibinfo  {journal} {Phys. Rev. B}\ }\textbf {\bibinfo {volume} {32}},\ \bibinfo {pages} {1778} (\bibinfo {year} {1985})}\BibitemShut {NoStop}%
\bibitem [{\citenamefont {Okumura}\ \emph {et~al.}(2011)\citenamefont {Okumura}, \citenamefont {Yoshino},\ and\ \citenamefont {Kawamura}}]{okumura_spin-chirality_2011}%
  \BibitemOpen
  \bibfield  {author} {\bibinfo {author} {\bibfnamefont {S.}~\bibnamefont {Okumura}}, \bibinfo {author} {\bibfnamefont {H.}~\bibnamefont {Yoshino}},\ and\ \bibinfo {author} {\bibfnamefont {H.}~\bibnamefont {Kawamura}},\ }\bibfield  {title} {\bibinfo {title} {Spin-chirality decoupling and critical properties of a two-dimensional fully frustrated {XY} model},\ }\href {https://doi.org/10.1103/PhysRevB.83.094429} {\bibfield  {journal} {\bibinfo  {journal} {Phys. Rev. B}\ }\textbf {\bibinfo {volume} {83}},\ \bibinfo {pages} {094429} (\bibinfo {year} {2011})}\BibitemShut {NoStop}%
\bibitem [{\citenamefont {Maghrebi}\ \emph {et~al.}(2017)\citenamefont {Maghrebi}, \citenamefont {Gong},\ and\ \citenamefont {Gorshkov}}]{maghrebi_continuous_2017}%
  \BibitemOpen
  \bibfield  {author} {\bibinfo {author} {\bibfnamefont {M.~F.}\ \bibnamefont {Maghrebi}}, \bibinfo {author} {\bibfnamefont {Z.-X.}\ \bibnamefont {Gong}},\ and\ \bibinfo {author} {\bibfnamefont {A.~V.}\ \bibnamefont {Gorshkov}},\ }\bibfield  {title} {\bibinfo {title} {Continuous symmetry breaking in {1D} long-range interacting quantum systems},\ }\href {https://doi.org/10.1103/PhysRevLett.119.023001} {\bibfield  {journal} {\bibinfo  {journal} {Phys. Rev. Lett.}\ }\textbf {\bibinfo {volume} {119}},\ \bibinfo {pages} {023001} (\bibinfo {year} {2017})}\BibitemShut {NoStop}%
\bibitem [{\citenamefont {Chen}\ \emph {et~al.}(2018)\citenamefont {Chen}, \citenamefont {Cheng}, \citenamefont {Xie}, \citenamefont {Wang},\ and\ \citenamefont {Xiang}}]{chen_equicalence_2018}%
  \BibitemOpen
  \bibfield  {author} {\bibinfo {author} {\bibfnamefont {J.}~\bibnamefont {Chen}}, \bibinfo {author} {\bibfnamefont {S.}~\bibnamefont {Cheng}}, \bibinfo {author} {\bibfnamefont {H.}~\bibnamefont {Xie}}, \bibinfo {author} {\bibfnamefont {L.}~\bibnamefont {Wang}},\ and\ \bibinfo {author} {\bibfnamefont {T.}~\bibnamefont {Xiang}},\ }\bibfield  {title} {\bibinfo {title} {{Equivalence of restricted Boltzmann machines and tensor network states}},\ }\href {https://doi.org/10.1103/PhysRevB.97.085104} {\bibfield  {journal} {\bibinfo  {journal} {Phys. Rev. B}\ }\textbf {\bibinfo {volume} {97}},\ \bibinfo {pages} {085104} (\bibinfo {year} {2018})}\BibitemShut {NoStop}%
\bibitem [{\citenamefont {Sneddon}(1995)}]{sneddon_fourier_1995}%
  \BibitemOpen
  \bibfield  {author} {\bibinfo {author} {\bibfnamefont {I.~N.}\ \bibnamefont {Sneddon}},\ }\href@noop {} {\emph {\bibinfo {title} {Fourier {Transforms}}}}\ (\bibinfo  {publisher} {Courier Corporation},\ \bibinfo {year} {1995})\BibitemShut {NoStop}%
\bibitem [{\citenamefont {Liu}\ \emph {et~al.}(2021)\citenamefont {Liu}, \citenamefont {Zhu}, \citenamefont {Huo}, \citenamefont {Feng}, \citenamefont {Song}, \citenamefont {Zhang}, \citenamefont {Chen}, \citenamefont {Lezec}, \citenamefont {Lu}, \citenamefont {Agrawal},\ and\ \citenamefont {Xu}}]{liu_multifunctional_2021}%
  \BibitemOpen
  \bibfield  {author} {\bibinfo {author} {\bibfnamefont {M.}~\bibnamefont {Liu}}, \bibinfo {author} {\bibfnamefont {W.}~\bibnamefont {Zhu}}, \bibinfo {author} {\bibfnamefont {P.}~\bibnamefont {Huo}}, \bibinfo {author} {\bibfnamefont {L.}~\bibnamefont {Feng}}, \bibinfo {author} {\bibfnamefont {M.}~\bibnamefont {Song}}, \bibinfo {author} {\bibfnamefont {C.}~\bibnamefont {Zhang}}, \bibinfo {author} {\bibfnamefont {L.}~\bibnamefont {Chen}}, \bibinfo {author} {\bibfnamefont {H.~J.}\ \bibnamefont {Lezec}}, \bibinfo {author} {\bibfnamefont {Y.}~\bibnamefont {Lu}}, \bibinfo {author} {\bibfnamefont {A.}~\bibnamefont {Agrawal}},\ and\ \bibinfo {author} {\bibfnamefont {T.}~\bibnamefont {Xu}},\ }\bibfield  {title} {\bibinfo {title} {Multifunctional metasurfaces enabled by simultaneous and independent control of phase and amplitude for orthogonal polarization states},\ }\href {https://doi.org/10.1038/s41377-021-00552-3} {\bibfield  {journal} {\bibinfo  {journal} {Light Sci. Appl.}\ }\textbf {\bibinfo {volume} {10}},\
  \bibinfo {pages} {107} (\bibinfo {year} {2021})}\BibitemShut {NoStop}%
\bibitem [{\citenamefont {Resnick}\ \emph {et~al.}(1981)\citenamefont {Resnick}, \citenamefont {Garland}, \citenamefont {Boyd}, \citenamefont {Shoemaker},\ and\ \citenamefont {Newrock}}]{resnick_kosterlitz-thouless_1981}%
  \BibitemOpen
  \bibfield  {author} {\bibinfo {author} {\bibfnamefont {D.~J.}\ \bibnamefont {Resnick}}, \bibinfo {author} {\bibfnamefont {J.~C.}\ \bibnamefont {Garland}}, \bibinfo {author} {\bibfnamefont {J.~T.}\ \bibnamefont {Boyd}}, \bibinfo {author} {\bibfnamefont {S.}~\bibnamefont {Shoemaker}},\ and\ \bibinfo {author} {\bibfnamefont {R.~S.}\ \bibnamefont {Newrock}},\ }\bibfield  {title} {\bibinfo {title} {Kosterlitz-{Thouless} transition in proximity-coupled superconducting arrays},\ }\href {https://doi.org/10.1103/PhysRevLett.47.1542} {\bibfield  {journal} {\bibinfo  {journal} {Phys. Rev. Lett.}\ }\textbf {\bibinfo {volume} {47}},\ \bibinfo {pages} {1542} (\bibinfo {year} {1981})}\BibitemShut {NoStop}%
\bibitem [{\citenamefont {Turtaev}\ \emph {et~al.}(2017)\citenamefont {Turtaev}, \citenamefont {Leite}, \citenamefont {Mitchell}, \citenamefont {Padgett}, \citenamefont {Phillips},\ and\ \citenamefont {Čižmár}}]{turtaev_comparison_2017}%
  \BibitemOpen
  \bibfield  {author} {\bibinfo {author} {\bibfnamefont {S.}~\bibnamefont {Turtaev}}, \bibinfo {author} {\bibfnamefont {I.~T.}\ \bibnamefont {Leite}}, \bibinfo {author} {\bibfnamefont {K.~J.}\ \bibnamefont {Mitchell}}, \bibinfo {author} {\bibfnamefont {M.~J.}\ \bibnamefont {Padgett}}, \bibinfo {author} {\bibfnamefont {D.~B.}\ \bibnamefont {Phillips}},\ and\ \bibinfo {author} {\bibfnamefont {T.}~\bibnamefont {Čižmár}},\ }\bibfield  {title} {\bibinfo {title} {Comparison of nematic liquid-crystal and {DMD} based spatial light modulation in complex photonics},\ }\href {https://doi.org/10.1364/OE.25.029874} {\bibfield  {journal} {\bibinfo  {journal} {Opt. Express}\ }\textbf {\bibinfo {volume} {25}},\ \bibinfo {pages} {29874} (\bibinfo {year} {2017})}\BibitemShut {NoStop}%
\bibitem [{\citenamefont {Yao}\ \emph {et~al.}(2021)\citenamefont {Yao}, \citenamefont {Lin},\ and\ \citenamefont {Bukov}}]{yao_reinforcement_2021}%
  \BibitemOpen
  \bibfield  {author} {\bibinfo {author} {\bibfnamefont {J.}~\bibnamefont {Yao}}, \bibinfo {author} {\bibfnamefont {L.}~\bibnamefont {Lin}},\ and\ \bibinfo {author} {\bibfnamefont {M.}~\bibnamefont {Bukov}},\ }\bibfield  {title} {\bibinfo {title} {Reinforcement learning for many-body ground-state preparation inspired by counterdiabatic driving},\ }\href {https://doi.org/10.1103/PhysRevX.11.031070} {\bibfield  {journal} {\bibinfo  {journal} {Phys. Rev. X}\ }\textbf {\bibinfo {volume} {11}},\ \bibinfo {pages} {031070} (\bibinfo {year} {2021})}\BibitemShut {NoStop}%
\bibitem [{\citenamefont {Mohseni}\ \emph {et~al.}(2022)\citenamefont {Mohseni}, \citenamefont {McMahon},\ and\ \citenamefont {Byrnes}}]{mohseni_ising_2022}%
  \BibitemOpen
  \bibfield  {author} {\bibinfo {author} {\bibfnamefont {N.}~\bibnamefont {Mohseni}}, \bibinfo {author} {\bibfnamefont {P.~L.}\ \bibnamefont {McMahon}},\ and\ \bibinfo {author} {\bibfnamefont {T.}~\bibnamefont {Byrnes}},\ }\bibfield  {title} {\bibinfo {title} {Ising machines as hardware solvers of combinatorial optimization problems},\ }\href {https://doi.org/10.1038/s42254-022-00440-8} {\bibfield  {journal} {\bibinfo  {journal} {Nat. Rev. Phys.}\ }\textbf {\bibinfo {volume} {4}},\ \bibinfo {pages} {363} (\bibinfo {year} {2022})}\BibitemShut {NoStop}%
\end{thebibliography}

%

\onecolumngrid
\appendix
\section{End Matter}
\twocolumngrid
\emph{XY spin encoding and Fourier mask}--The SLM and camera are placed on the front and back focal planes of a lens such that the lens performs an optical Fourier transform for the modulated field. The distribution of the intensity on the camera can be described as 
\begin{equation}
I(\boldsymbol{u})=\left|\mathcal{F}\left\{\sum_{j=1}^N A_j e^{i\varphi_j} \operatorname{rect}\left(\frac{\boldsymbol{x}-\boldsymbol{x}_j}{W}\right)\right\}\right|^2 
\end{equation}
where $\mathcal{F}$ means the Fourier transform, $A_j$ and $\varphi_j$ are the amplitude and phase of modulated light for the $j$-th spin, $\boldsymbol{x}=(x,~y)$ and $\boldsymbol{u}=(u, v)$ are the coordinates on the SLM and its Fourier plane of the lens, $\boldsymbol{x}_j$ is the coordinate of the pixel center representing the $j$-th spin, $W$ is the length of the square pixels allocated to each spin on the SLM, and $\mathrm{rect}(\boldsymbol{x})$ is the rectangular function which takes 1 if $|x|\leq 1/2$ and $|y|\leq1/2$, otherwise 0. The phase $\varphi_j=\theta_j$ and the amplitude $A_j$ is proportional to the disorder strength $\xi_j$ in Eq. (\ref{equ1}) (see \cite{supp} for more details).\par

The Hamiltonian of the XY model is obtained by making integration of $I(\bm{u})I_M(\bm{u})$, where $I_M(\bm{u})$ is a Fourier mask \cite{fan_programmable_2023} designed for specific spin interactions. According to the convolution theorem \cite{sneddon_fourier_1995}, the integral of light intensity $I(\boldsymbol{u})$ multiplied by a modulation function $I_M(\boldsymbol{u})$ is equivalent to performing the convolution for all spins with the kernel $\mathcal{F}^{-1}\left\{I_M(\boldsymbol{u})\right\}$ \cite{supp}. When $\mathcal{F}^{-1}\left\{I_M(\boldsymbol{u})\right\}$ takes the form of summation over several delta functions, representing the interaction between a spin to others at discrete lattice sites, the corresponding Fourier mask is 
\begin{equation}
I_M(\boldsymbol{u})=\mathcal{F}\left\{\sum_{\alpha,\boldsymbol{r}^{(\alpha)}}J_\alpha\delta\left(\boldsymbol{x}-\boldsymbol{r}^{(\alpha)}\right) \right\}~,
\label{equ2}
\end{equation}
where $\boldsymbol{r}^{(\alpha)}$ is the distance between interacting spin sites. $J_{\alpha}$ controls the ratio of various Fourier masks by weighted sum and the summation over $\alpha$ indicates that arbitrary range interactions can be achieved by a Fourier mask. Therefore, the Hamiltonian of the XY model can be directly obtained in experiment by a correspondingly designed Fourier mask, which accelerates the sampling of spin configurations (see derivation in \cite{supp}). In our experiment, the Fourier mask is linearly renormalized to values between 0 and 1, which allows experimental implementation with an intensity attenuator such as a metasurface \cite{liu_multifunctional_2021}.

\emph{Computational complexity of the FPS}--Due to geometrical frustration, the energy landscape of the XY model is generally rough such that it takes a long time to obtain a sufficient number of samples for large-scale lattices. FPS directly obtains the Hamiltonian of spin configurations from the measured light intensity and updates it using MCMC algorithms, which significantly speeds up the sampling process. By incorporating the Fourier-mask technique into the FPS, we can employ a single Fourier mask to realize arbitrary range spin interactions. Consequently, the speed of calculating the Hamiltonian is nearly the same for any range of spin interactions, and the computational complexity is $\mathcal{O}(N)$, while most of the current local updating algorithms scale as $\mathcal{O}(N\mathrm{log}N)$ \cite{muller_fast_2023} ({see \cite{supp} for detailed information}). This feature enables efficient calculation of large-scale XY models with complex long-range interactions \cite{giachetti_berezinskii-kosterlitz-thouless_2021}, and can be further extended to simulate statistical models that can be mapped to XY models, including superfluids \cite{kosterlitz_long_1972}, superconductors \cite{kosterlitz_nobel_2017}, and Josephson junctions arrays \cite{resnick_kosterlitz-thouless_1981}. We also provide some results on long-range interactions featuring power-law decay in square lattices (see \cite{supp}).\par

The time consumption of FPS includes the response time of the SLM, the acquisition time of the camera, and the computation time of Hadamard multiplication and summation between $I$ and $I_M$. In our implementation of the FPS, the SLM is replaced by a digital micromirror device, which has a refresh rate of 20~kHz, significantly increasing the modulation speed of the FPS \cite{turtaev_comparison_2017}. However, this step also introduces a large optical aberration caused by the industrial-grade chip of digital light processing, resulting in significant energy deviations. To address this issue, we develop an iterative retrieval approach to precompensate the overall aberration of FPS (see \cite{supp} for the algorithm details and the results of the aberration compensation), enabling to precisely obtain the Hamiltonian of the XY models. For further acceleration, a metasurface \cite{liu_multifunctional_2021} and photodiode can be used to replace the camera to reduce the overall time consumption. We can further speed up the sampling procedure by using state-of-the-art machine learning algorithms to suggest optimized spin-flip policies \cite{yao_reinforcement_2021}, which can be used to solve optimization problems in real time \cite{mohseni_ising_2022}.

\end{document}



\title{Supplementary Materials for: Topological Defects and Geometrical Frustrations in Fourier Photonic
Simulator}






\affiliation{Zhejiang Key Laboratory of Micro-Nano Quantum Chips and Quantum Control, School of Physics, and State Key Laboratory for
Extreme Photonics and Instrumentation, Zhejiang University, Hangzhou 310027, China} 
\affiliation{College of Optical Science and Engineering, Zhejiang University, Hangzhou 310027, Zhejiang Province, China}
\affiliation{Hefei National Laboratory, Hefei 230088, China}
\affiliation{Institute for Advanced Study in Physics and School of Physics, Zhejiang University, Hangzhou, 310058, China}

\author{Yuxuan Sun$^{1,*}$}
\author{Weiru Fan$^{1,*,\dagger}$}
\author{Xingqi Xu$^{1}$}
\author{Da-Wei Wang$^{1,2,3,\ddagger}$}
\author{Hai-Qing Lin$^{4,\S}$}





\maketitle

\tableofcontents

\section{S1. The Fourier-mask enabled Photonic Simulator}\label{sec_s1}
We employ the gauge transformation proposed in Ref. \cite{fang_experimental_2021} to encode both disordered spin interaction strengths $\xi_j$ and XY spins $\boldsymbol{S}_j$ into effective XY spins $\boldsymbol{S}_j^\prime$ utilizing a single spatial light modulator (SLM). These effective XY spins are applied to individual regions on the SLM with the $\varphi_j=\theta_j+\arccos\xi_j$, where $\theta_j$ is the angle of the spin $\boldsymbol{S}_j$ and $\xi_j$ is the disorder strength. Therefore, the modulated input field can be expressed as
\begin{equation}
E(\boldsymbol{x})=\sum_{j=1}^N A_j e^{i\varphi_j} \operatorname{rect}\left(\frac{\boldsymbol{x}-\boldsymbol{x}_j}{W}\right)=E_0\left[ \sum_{j=1}^N \xi_j e^{i\theta_j} \delta\left(\boldsymbol{x}-\boldsymbol{x}_j\right)\right] * \operatorname{rect}\left(\frac{\boldsymbol{x}}{W}\right),
\label{equ_s1}
\end{equation}
where $E_0$ is a scalar factor and the amplitude $A_j = E_0\xi_j$. $\boldsymbol{x}=(x,y)$ is the spatial coordinate on the SLM (real space), $\boldsymbol{x}_j$ is the position of the $j$-th pixel and $W$ is the pixel size allocated to each spin on the SLM. $*$ represents the convolution operation. The rectangular function is defined as,
\begin{equation}
\operatorname{rect}\left(\frac{\boldsymbol{x}}{W}\right)= \begin{cases}1 & |x|,|y| \leq W / 2 \\ 0 & |x|,|y|>W / 2\end{cases}.
\label{equ_s2}
\end{equation}

According to the theory of Fourier optics \cite{goodman_introduction_2005}, the modulated light field located on the front focal plane of a lens can perform optical Fourier transformation (FT), and its result is observed on the back focal plane. When a camera is placed on the back focal plane, the recorded intensity distribution is
\begin{equation}
I(\boldsymbol{u})=I_0 \sum_{j, h=1}^N \xi_j \xi_h e^{i \left(\theta_j-\theta_h\right)} \exp \left(i \frac{2 \pi}{f \lambda}\boldsymbol{r}_{jh} \cdot \boldsymbol{u}\right) \operatorname{sinc}^2\left(\frac{\boldsymbol{u} W}{f \lambda}\right),
\label{equ_s3}
\end{equation}
where $I_0$ is the global intensity factor, $f$ is the focal length of the lens, $\lambda$ is the wavelength and $\boldsymbol{u}=(u,v)$ is the spatial coordinate on the camera (Fourier space). $\boldsymbol{r}_{jh}=\boldsymbol{x}_j-\boldsymbol{x}_h$ is the distance vector between the $j$-th and $h$-th spins. When the recorded intensity distribution is further modulated by a spatial intensity attenuator $I_M(\boldsymbol{u})$, the finally detected signal can be expressed as
\begin{equation}
\begin{aligned}
 \int I(\boldsymbol{u}) I_M(\boldsymbol{u}) d \boldsymbol{u} \approx & I_0 \sum_{j, h=1}^N \xi_j \xi_h e^{i \left(\theta_j-\theta_h\right)} \int I_M(\boldsymbol{u}) \exp \left(i \frac{2 \pi}{f \lambda}\boldsymbol{r}_{jh} \cdot \boldsymbol{u}\right) d\boldsymbol{u}\,,
\end{aligned}
\label{equ_s4}
\end{equation}
where the size effect induced by the finite $W$ is negligible such that $\operatorname{sinc}^2\left(\frac{\boldsymbol{u} W}{f \lambda}\right)\sim 1$. By applying the convolution theorem, the Eq. \ref{equ_s4} can be further rewritten as
\begin{equation}
\begin{aligned}
    \int I(\boldsymbol{u}) I_M(\boldsymbol{u}) d \boldsymbol{u} 
= & I_0 \sum_{j, h=1}^N \xi_j \xi_h e^{i \left(\theta_j-\theta_h\right)} \int \mathcal{F}\left\{\mathcal{F}^{-1}\left\{I_M(\boldsymbol{u}) \right\} * \mathcal{F}^{-1}\left\{\exp \left(i \frac{2 \pi}{f \lambda}\boldsymbol{r}_{jh} \cdot \boldsymbol{u}\right)\right\}\right\}d\boldsymbol{u}\\
= & I_0 \sum_{j, h=1}^N \xi_j \xi_h e^{i \left(\theta_j-\theta_h\right)} \int \mathcal{F}\left\{\mathcal{F}^{-1}\left\{I_M(\boldsymbol{u}) \right\} * \delta\left(\boldsymbol{\nu}+\frac{\boldsymbol{r}_{jh}}{f\lambda}\right) \right\}d\boldsymbol{u}\,.
\label{equ_s4_1}
\end{aligned}
\end{equation}
Here, $\boldsymbol{\nu}$ is the spatial frequency corresponding to $\boldsymbol{u}$. We set $\mathcal{F}^{-1}\left\{I_M(\boldsymbol{u}) \right\}$ as the summation of delta functions,
\begin{equation}
\mathcal{F}^{-1}\left\{I_M(\boldsymbol{u})\right\} = \sum_{\alpha\in \mathbb{Z}^+,\boldsymbol{r}^{(\alpha)}}J_\alpha\delta\left(\boldsymbol{\nu}-\frac{\boldsymbol{r}^{(\alpha)}}{f\lambda}\right) \,,
\label{equ_s4_2}
\end{equation}
where $\alpha$ in the summation designates spin pairs, such as 1 for nearest-neighbor (NN) spin pairs and 2 for next-nearest-neighbor (NNN) spin pairs, $\boldsymbol{r}^{(\alpha)}$ is the
distance between coupling spin sites, and $J_{\alpha}$ is the interaction strength for different ranges of neighbors. Substituting Eq. \ref{equ_s4_2} into Eq. \ref{equ_s4_1}, we derive
\begin{equation}
\begin{aligned}
    \int I(\boldsymbol{u}) I_M(\boldsymbol{u}) d \boldsymbol{u}
= & I_0 \sum_{j, h=1}^N \xi_j \xi_h e^{i \left(\theta_j-\theta_h\right)}  \sum_{\alpha\in \mathbb{Z}^+,\boldsymbol{r}^{(\alpha)}}J_\alpha \int \mathcal{F}\left\{\delta\left(\boldsymbol{\nu}-\frac{\boldsymbol{r}^{(\alpha)}}{f\lambda}\right) * \delta\left(\boldsymbol{\nu}+\frac{\boldsymbol{r}_{jh}}{f\lambda}\right) \right\}d\boldsymbol{u} \\
= & I_0 \sum_{j, h=1}^N \xi_j \xi_h e^{i \left(\theta_j-\theta_h\right)}  \sum_{\alpha\in \mathbb{Z}^+,\boldsymbol{r}^{(\alpha)}}J_\alpha\delta\left(\frac{2\pi}{f\lambda}\left( \boldsymbol{r}_{jh}-\boldsymbol{r}^{(\alpha)}\right)\right)\\
= & 2I_0 \sum_{\alpha\in \mathbb{Z}^+} J_\alpha\sum_{(j,\,h)_\alpha}\xi_j\xi_h\cos\left(\theta_j-\theta_h\right)\,.
\label{equ_s4_3}
\end{aligned}
\end{equation}
Eq. \ref{equ_s4_3} demonstrates the direct relationship between the optical propagation system and XY model Hamiltonian (Eq.~1 in the main text). Because $I_M(\boldsymbol{u})$ modulates the intensity in Fourier space, we refer to it as the Fourier mask and it can be designed according to Eq. \ref{equ_s4_2}.

For NN interaction on the square lattice, the corresponding Fourier mask is 
\begin{equation}
    \begin{aligned}
        I_M(\boldsymbol{u}) = & \mathcal{F}\left\{\left[\delta\left(\boldsymbol{\nu}\pm \frac{W\hat{\boldsymbol{e}}_x}{f\lambda}\right)+\delta\left(\boldsymbol{\nu}\pm \frac{W\hat{\boldsymbol{e}}_y}{f\lambda}\right)\right]\right\}\\
        = & 2\mathrm{cos}\left(\frac{2\pi W}{f\lambda}u\right) +2\mathrm{cos}\left(\frac{2\pi W}{f\lambda}v\right)\,,
    \end{aligned}
\end{equation}
with values ranging from -4 to 4. To facilitate the implementation of the Fourier mask on optical apparatus, we shift and renormalize $I_M(\boldsymbol{u})$ to values between 0 and 1 as $[\mathrm{cos}(\frac{2\pi W}{f\lambda}u) +\mathrm{cos}(\frac{2\pi W}{f\lambda}v)+2]/{4}$. This enables the future implementation of Fourier mask with a metasurface or an intensity attenuator. Similarly, the renormalized $I_M$ for the NNN interaction is $[\mathrm{cos}(\frac{2\pi W}{f\lambda}u) \cdot\mathrm{cos}(\frac{2\pi W}{f\lambda}v)+1]/{2}$.

However, the procedure of renormalization introduces additional non-zero self couplings, $J_0$. Hence, the Eq. \ref{equ_s4_3} becomes
\begin{equation}
    \int I(\boldsymbol{u}) I_M(\boldsymbol{u}) d \boldsymbol{u} = 2I_0 \sum_{\alpha\in \mathbb{Z}^+} J_\alpha\sum_{(j,\,h)_\alpha}\xi_j\xi_h\cos\left(\theta_j-\theta_h\right)+I_0J_0N\,.
    \label{equ_s5}
\end{equation}
Moreover, the Eq. \ref{equ_s5} exhibits discrepancies with the Hamiltonian of XY models (Eq. 1 in the main text) with an intensity factor. To precisely obtain the Hamiltonian, we employ the unbiased normalization approach \cite{fan_programmable_2023}, eliminating self couplings and global intensity factor $2I_0$. Specifically, the initial step is to set all spins in the same direction, resulting in the intensity distribution $I_{\text{init}}(\boldsymbol{u})$. The Hamiltonian for this specific spin configuration, denoted as $\mathcal{H}_{\text{init}}(\boldsymbol{u})$, can be easily calculated. For example, it is $-2N$ for the 2D XY model with NN interaction on the square lattice. For the self spin couplings term in Eq. (\ref{equ_s4}), it can be obtained by selecting a spin configuration with a zero Hamiltonian, such as a stripe pattern for NN interaction. The intensity distribution of this configuration, corresponding to zero Hamiltonian, is recorded as $I_{\text{cali}}(\boldsymbol{u})$. Finally, the Hamiltonian for an arbitrary spin configuration is expressed as
\begin{equation}
\mathcal{H}=\frac{\mathcal{H}_{\text {init }}}{\mathcal{I}_{\text{init}}-\mathcal{I}_{\text{cali}}} \left(\int I_M(\boldsymbol{u}) I(\boldsymbol{u}) d \boldsymbol{u}-\mathcal{I}_{\text{cali}}\right),
\label{equ_s6}
\end{equation}
where $\mathcal{I}_{\text{init}}=\int I_M(\boldsymbol{u}) I_{\text {init }}(\boldsymbol{u}) d \boldsymbol{u}$ and $\mathcal{I}_{\text{cali}}=\int I_M(\boldsymbol{u}) I_{\text {cali }}(\boldsymbol{u}) d \boldsymbol{u}$.

More generally, the gauge transformation is unnecessary when we use complex field modulation. In this case, the wavefront amplitude and phase of light field can be simultaneously manipulated, realizing arbitrary XY spins and disordered spin interaction strengths. While the formalism of the Fourier mask is identical, owing to the consistency of the above analysis.

\section{S2. Spin Flipping and Sampling Algorithm}
We use the Metropolis-Hastings spin flipping algorithm to perform simulated annealing and update the spin configuration. In the experiment, an initial spin configuration is generated by random sampling from uniform distribution, which is eligible at high temperatures. Loading it into our FPS, the corresponding Hamiltonian can be obtained. Then, a site is randomly selected and the spin on it is flipped, after which its Hamiltonian is obtained by FPS again. The energy difference $\Delta\mathcal{H}$ can be computed by the Hamiltonians before and after the flip. This flip is accepted with a probability ($P_{\text{acc}}$) as determined by the Metropolis-Hastings rule $P_{\text{acc}}=\min\left\{1,~\exp (-\beta \Delta \mathcal{H})\right\}$, where $\beta = 1/k_BT$ is the inverse temperature and $k_B$ is the Boltzmann constant. Traditionally, only one spin is flipped at a time, leading to small energy changes that are susceptible to experimental noise. To overcome this, we generalize the single-spin-flipping to the multiple-spin-flipping while still applying the same acceptance rule. To accelerate sampling, the flipping angle is gradually diminished with the temperature, which obeys an exponential decay. With the implementation of spin flipping, the system will eventually evolve into a stable state at current temperature $T$. 
\par

In order to estimate physical quantities of XY models, we obtain $n$ configuration samples by sampling from the stable state under each effective temperature $T$, denoted as $\left\{\left\{s_{T,1}\right\},\left\{s_{T,2}\right\},\cdots,\left\{s_{T,n}\right\}\right\}$, which obey the Boltzmann distribution. Here, these samples are obtained by flipping single spin with a fixed step, ensuring that these samples are independent identically distributed for effective ensemble estimation. Once sufficient configuration samples are generated, we can calculate the average magnetization $m$, magnetic susceptibility $\chi$, helical modulus $Y$, and heat capacity $C_V$ by the equations at main text. By studying these quantities, the dynamics of annealing can be observed, such as phase transitions, frustration and ground state properties.\par 

\section{S3. Experimental Setup and Phase Modulation}\label{sec_s3}
\begin{figure}[hbt]
    \centering
    \includegraphics[width=0.5\linewidth]{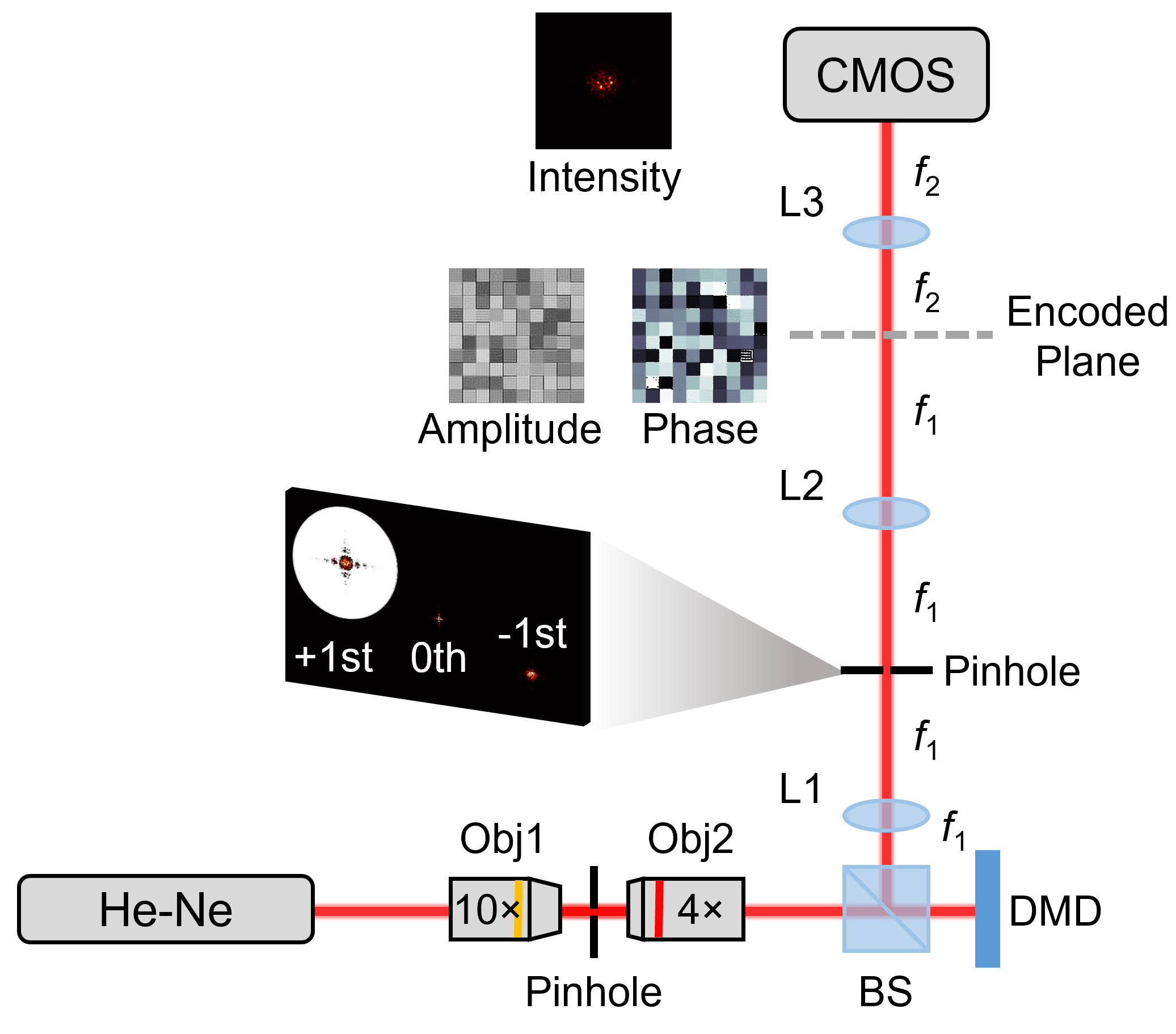}
    \caption{\label{fig_s1}Experimental setup. The phase and amplitude patterns were generated at the encoded plane by the superpixel method. A pinhole is used to select out the first-order of diffraction light. Obj1 and Obj2: objectives; BS: beam splitter; DMD: digital micro-mirror device; L$_1$, L$_2$ and L$_3$: lenses; CMOS: complementary metal-oxide-semiconductor sensor.}
\end{figure}
\begin{figure}[hbt]
    \centering
    \includegraphics[width=0.75\linewidth]{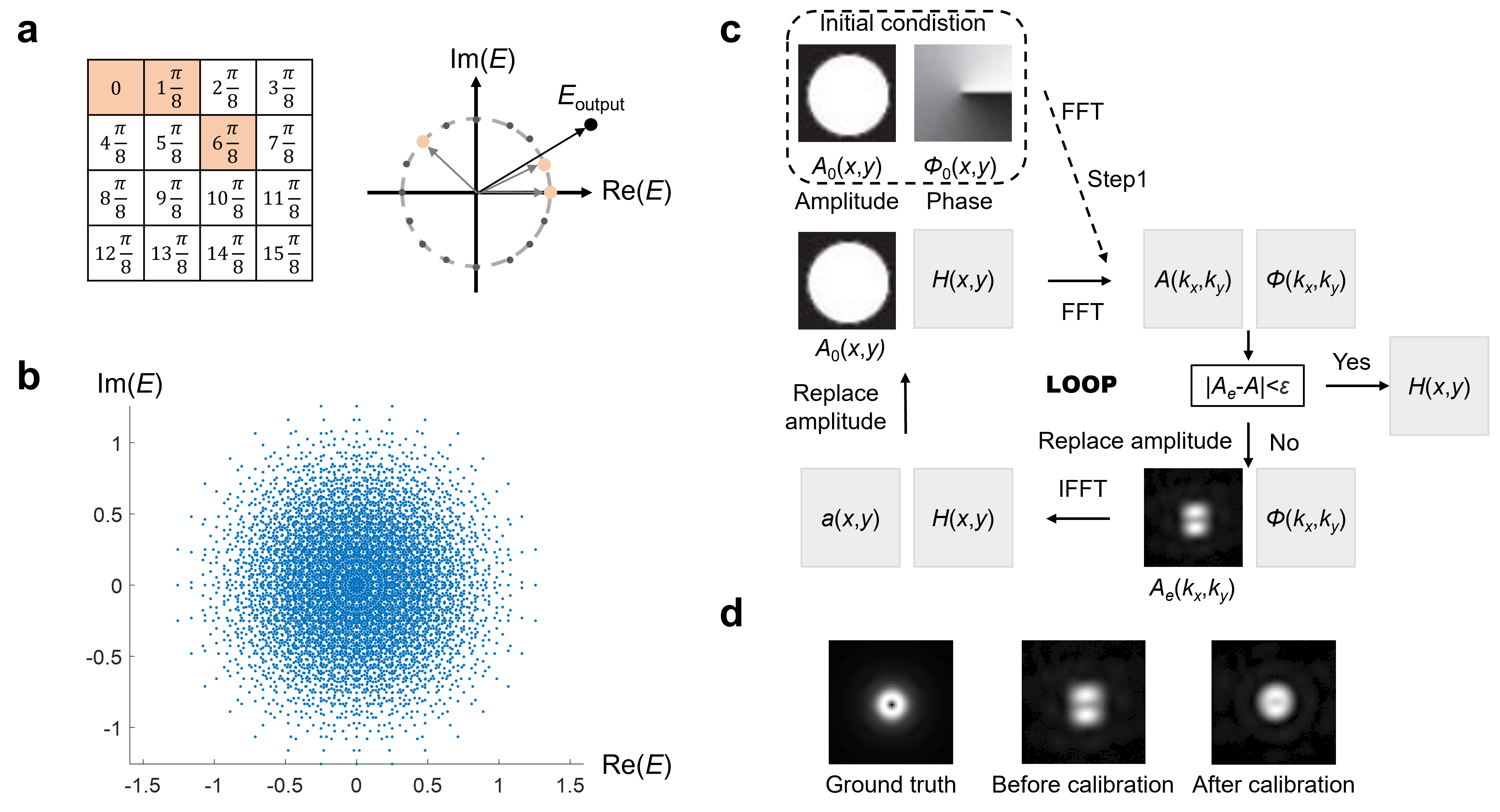}
    \caption{\label{fig_s1p} Complex amplitude modulation and the wavefront compensation. (a) The working principle of superpixel method \cite{Goorden_superpixel-based_2014}. Yellow and white regions represent the on and off state of DMD {pixels}, respectively. The value of each region is the angle of the unit vector in the right subgraph. (b) All complex amplitude that can be achieved by the superpixel method. (c) The workflow of GS algorithm for phase retrieval\cite{jesacher_wavefront_2007}. (d) The ground truth generated by simulation, and the experimental intensity pattern before and after compensation.}
\end{figure}

The experimental setup is shown in FIG.~\ref{fig_s1}. The coherent light was generated by a He-Ne laser (LASOS, LGK 7634), and shaped by a a spatial filter consisting of a pair of objectives (Olympus, 10$\times$, NA=0.3 and Olympus, 4$\times$, NA=0.13) and a pinhole (Thorlabs, P20K, $\Phi=$ 20 $\mathrm{\mu m}$) to obtain a quasi-plane wave with 8 $mm$ diameter. Then, the light passed through a beam splitter (BS) and was modulated by a digital micro-mirror device (DMD; VIALUX, V-9501). Subsequently, the modulated laser light was focused by lens L$_1$ (focal length 100mm, $f_1$), and its first-order diffraction light was filtered out (see the insert in FIG.~\ref{fig_s1}). Lens L$_2$ (focal length 100mm, $f_1$) was used to transform the diffraction light from the pinhole plane to encoded plane, where we can realize the complex amplitude modulation. The modulated beam was performed a Fourier transform by L3 ($f_2$), and was recorded by a CMOS camera (TUCSEN, Dhyana 400BSI) at the focal plane. In our experiment, the complex amplitude modulation is achieved by the superpixel method \cite{Goorden_superpixel-based_2014}, where adjacent 4 $\times$ 4 micromirrors are grouped to form a superpixel. The output complex field of a superpixel was controlled by the on-off state of the 16 micromirrors, as shown in FIG.~\ref{fig_s1p}a, and all the possible output complex fields are shown in FIG.~\ref{fig_s1p}b. The on-off state of each pixel is determined by referring to a lookup table according to the real and imaginary parts of the desired complex amplitude.\par

Traditionally, DMDs are not optically flat due to imperfections in the manufacturing process. This introduces an abysmal phase background, leading to a variable energy bias during the annealing and sampling process. Therefore, simulations are severely affected and even unavailable. To address this, we use the phase iteration algorithm to calculate the phase background, then correct the wavefront distortion introduced by DMD. Here, we use the Gerchberg-Saxon (GS) algorithm to perform phase retrieval \cite{jesacher_wavefront_2007}. Specifically, we encode a helical phase and a circular aperture onto the wavefront, and perform fast Fourier transform (FFT) to generate the amplitude and phase in Fourier plane, as well as obtain the real amplitude on the back focal plane of the lens in the experiment. After that, we replace the simulated amplitude with the experimental amplitude and perform the inverse fast Fourier transform (IFFT) to generate a new conjectural amplitude and phase. We sequentially replace the conjectural amplitude with the circular aperture, and repeat these procedures until convergence or satisfying the stopping condition. Finally, the output retrieved phase is the distortion phase superimposed on the helical phase. The operation of the GS algorithm is sketched in FIG.~\ref{fig_s1p}c, and the intensity patterns on the back focal plane before and after compensation are shown in FIG.~\ref{fig_s1p}d. \par

\begin{figure}[h]
\centering
\includegraphics[width=0.8\linewidth]{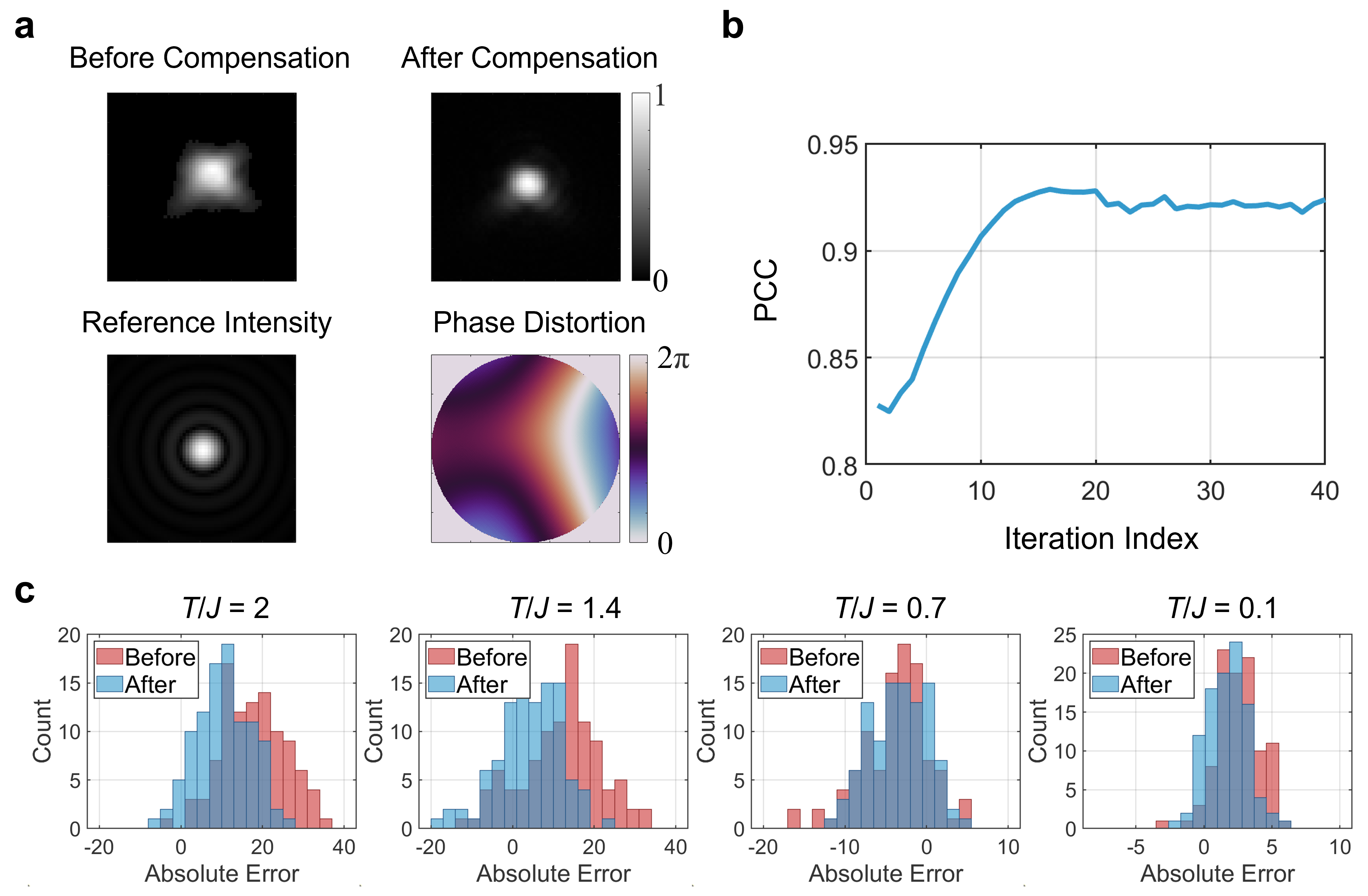}
\caption{\label{fig_r1}Quantitative results of optical aberration compensation. (a) Intensity distribution before and after aberration compensation, ideal reference intensity without aberration, and the estimated optical aberration. (b) The Pearson correlation coefficient (PCC) between the compensated and referenced intensity during the iterative retrieval. (c) The distribution of the energy error. The mean $\pm$ standard deviation before (after) compensation are $17.8 \pm 7.9$ ($10.4 \pm 6.7$), $12.1 \pm 9.5$ ($4.4 \pm 7.9$), $-3.8 \pm 4.0$ ($-3.3 \pm 3.5$), and $2.5 \pm 1.6$ ($1.7 \pm 1.4$) for $T/J = $ 2, 1.4, 0.7 and 0.1, respectively.}
\end{figure}

We also exhibit the results of compensation and estimated optical aberration. In FIG.~\ref{fig_r1}a, we show the compensated results and estimated optical aberration. Specifically, aberrations introduced by the DMD cause an irregular distortion of the ideal Gaussian beam. After the iterative compensation, the spot returns to a Gaussian distribution. The quantitative similarity factor (PCC) is calculated during the procedure of phase compensation (FIG.~\ref{fig_r1}b). As a result, phase compensation effectively reduces Hamiltonian calculation errors in the experiment (FIG.~\ref{fig_r1}c). \par

\section{S4. Tuning the Lattice Structures and Spin Interactions}

\begin{figure}[h]
\centering
\includegraphics[width=0.7\linewidth]{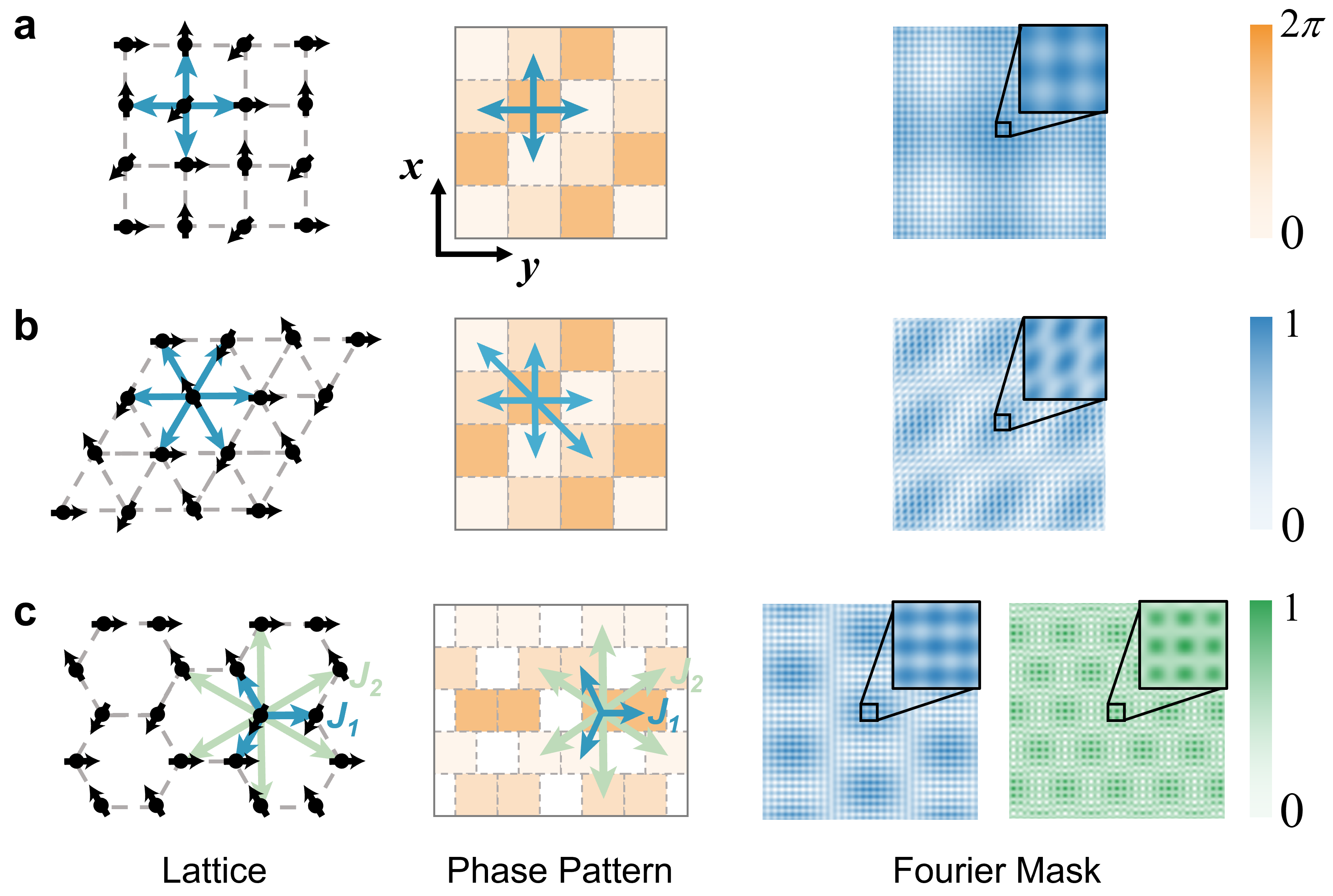}
\caption{\label{fig_s2}Synthesizing different lattice structures and spin interactions. (a) Ferromagnetic XY model on the square lattice. (b) Antiferromagnetic XY model on the triangular lattice. (c) Antiferromagnetic $J_1$-$J_2$ XY model on the honeycomb lattice. From the left column to the right column are the original lattice structures, the encoding pixel distribution on the DMD, and the Fourier masks to synthesize corresponding interactions, respectively.}
\end{figure}

\textit{Triangular Lattice.} The pixels of an SLM are generally square, such that the square lattice can be straightforward constructed. In this case, XY spins are successively encoded on individual areas of the DMD, as shown in FIG.~\ref{fig_s2}a. For the triangular lattice, one categorical approach is to combine multiple square pixels into a single hexagon area, enabling to encode one XY spin. A drawback of this method is the unsmoothness (zigzag shape) on the edges, causing a deviation in the obtained Hamiltonian. Moreover, grouping multiple pixels also deplete the resources of SLM. To overcome this, we rethink the Fourier mask and the degree of freedom on wavefront. Compared with the square lattice, the triangular lattice can be conceptualized as a stretched version of the square lattice, with each row displaced relative to the adjacent row. Parallelly, the Fourier-mask can realize any translation invariant spin interactions. Therefore, we can construct a triangular lattice with square SLM's pixel by meticulously designing the matching Fourier mask. For XY model on the triangular lattice with the NN spin interactions, the matching Fourier-mask involves NN interactions and additional NNN interactions along the main diagonal on original square lattice, as shown in the middle column in FIG.~\ref{fig_s2}b. Such Fourier-mask can be calculated by
\begin{equation}
    \begin{aligned}
        I_M^{tri}(\boldsymbol{u}) = & \mathcal{F}\left\{\left[\delta\left(\boldsymbol{\nu}\pm \frac{W\hat{\boldsymbol{e}}_x}{f\lambda}\right)+\delta\left(\boldsymbol{\nu}\pm \frac{W\hat{\boldsymbol{e}}_y}{f\lambda}\right)+\delta\left(\boldsymbol{\nu}\pm \frac{W\hat{\boldsymbol{e}}_x-W\hat{\boldsymbol{e}}_y}{f\lambda}\right)\right]\right\}\\
        = & 2\mathrm{cos}\left(\frac{2\pi W}{f\lambda}u\right) + 2\mathrm{cos}\left(\frac{2\pi W}{f\lambda}v\right) + 2\mathrm{cos}\left(\frac{2\pi W}{f\lambda}(u-v)\right)\,,
    \end{aligned}
\end{equation}
corresponding Fourier mask is shown in the right panel of FIG.~\ref{fig_s2}b.

\textit{Honeycomb Lattice.} When the honeycomb lattice is squeezed and stretched to a square lattice, vacancies appear at specific positions with fixed displacements. These vacancies are easily to realize by modulating zero light intensity onto the corresponding areas of the wavefront. For simplicity, these vacancies are referred to as empty areas and others are referred to as encoded areas. As shown in the middle panel of FIG.~\ref{fig_s2}c, there is an empty area after every two encoded areas in each row, with a displacement between adjacent rows. In this case, the matching Fourier masks for $J_1$ and $J_2$ are 
\begin{equation}
\label{equ_s6_1}
    \begin{aligned}
        I_M^{hon,\,J_1}(\boldsymbol{u}) = & \mathcal{F}\left\{\left[\delta\left(\boldsymbol{\nu}\pm \frac{W\hat{\boldsymbol{e}}_x}{f\lambda}\right)+\delta\left(\boldsymbol{\nu}\pm \frac{W/2\,\hat{\boldsymbol{e}}_x-W\hat{\boldsymbol{e}}_y}{f\lambda}\right)+\delta\left(\boldsymbol{\nu}\pm \frac{W/2\,\hat{\boldsymbol{e}}_x+W\hat{\boldsymbol{e}}_y}{f\lambda}\right)\right]\right\}\\
        = & 2\mathrm{cos}\left(\frac{2\pi W}{f\lambda}u\right)+2\mathrm{cos}\left(\frac{2\pi W}{f\lambda}(u/2-v)\right)+2\mathrm{cos}\left(\frac{2\pi W}{f\lambda}(u/2+v)\right)
    \end{aligned}
\end{equation}
\begin{equation}
\label{equ_s6_2}
    \begin{aligned}
        I_M^{hon,\,J_2}(\boldsymbol{u}) = & \mathcal{F}\left\{\left[\delta\left(\boldsymbol{\nu}\pm \frac{W\hat{\boldsymbol{e}}_y}{f\lambda}\right)+\delta\left(\boldsymbol{\nu}\pm \frac{3W/2\,\hat{\boldsymbol{e}}_x-W\hat{\boldsymbol{e}}_y}{f\lambda}\right)+\delta\left(\boldsymbol{\nu}\pm \frac{3W/2\,\hat{\boldsymbol{e}}_x+W\hat{\boldsymbol{e}}_y}{f\lambda}\right)\right]\right\}\\
        = & 2\mathrm{cos}\left(\frac{2\pi W}{f\lambda}v\right)+2\mathrm{cos}\left(\frac{2\pi W}{f\lambda}(3u/2-v)\right)+2\mathrm{cos}\left(\frac{2\pi W}{f\lambda}(3u/2+v)\right)
    \end{aligned}
\end{equation}
where Eqs. \ref{equ_s6_1} and \ref{equ_s6_2} correspond to the blue and green figures in the right column of FIG.~\ref{fig_s2}c, respectively. Different ratios $J_1/J_2$ can be realized by weighted summation of the two Fourier masks with weight of $J_1/J_2$.

\section{S5. Ground States with Geometrical Frustration}
\begin{figure}[hbt]
\centering
\includegraphics[width=0.4\linewidth]{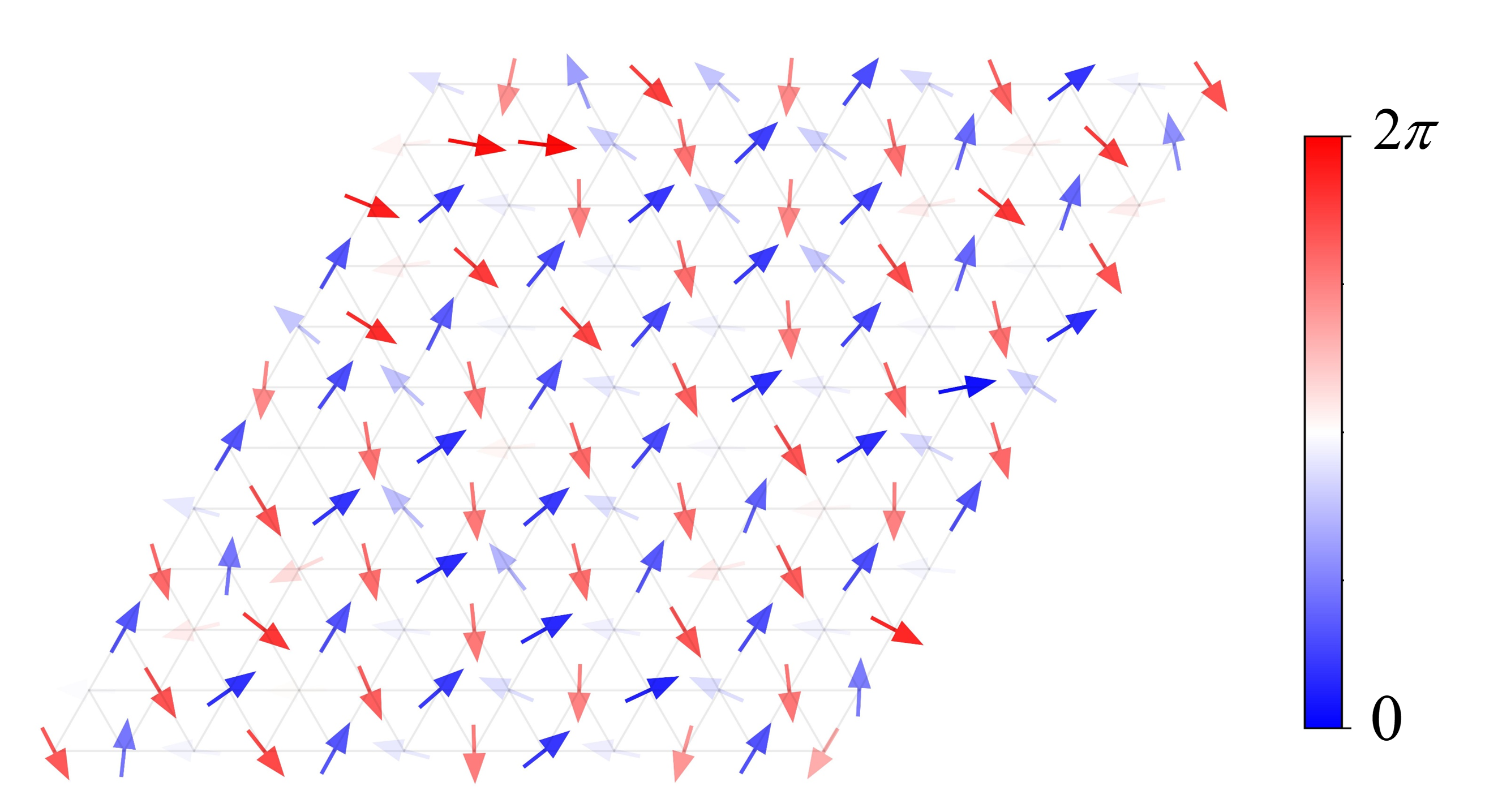}
\caption{\label{fig_s3}The low-temperature state generated by FPS of XY models on the triangular lattice with NN and antiferromagnetic spin interactions at $T/J=0.25$. The angle between any two spin pairs is $120^\circ$, indicating the $120^\circ$ order.}
\end{figure}

Geometrical frustration, arising from the conflicting interactions between spins, leads to extremely complex spin configurations and brings novel phases and orders \cite{bergman_order-by-disorder_2007}. For instance, in a triangular lattice with antiferromagnetic spin interactions, nearest-neighbor spin pairs tend to align antiparallel due to the antiferromagnetic coupling. However, the spins the at three vertices of each triangular unit cannot simultaneously align antiparallel to each other. As a compromised result, the angle between any two spin pairs turns out to be $120^\circ$ within each triangular unit at low temperature, which is distinct from traditional antiferromagnetic phase where all spin pairs are antiparallel. To study this phenomenon, we implement the search of
ground state by FPS, and the result is shown in FIG.~\ref{fig_s3}. The result exhibits the $120^\circ$ order, verifying the feasibility of using FPS to directly observe geometrical frustration. It is worth noting that the exact ground states can only be observed at absolute zero temperature and our experiment set ultimate temperature to $T/J=0.25$. Therefore, the angle between any two spin pairs occasionally deviates slightly from the ideal $120^\circ$.

\begin{figure}[hbt]
\centering
\includegraphics[width=0.7\linewidth]{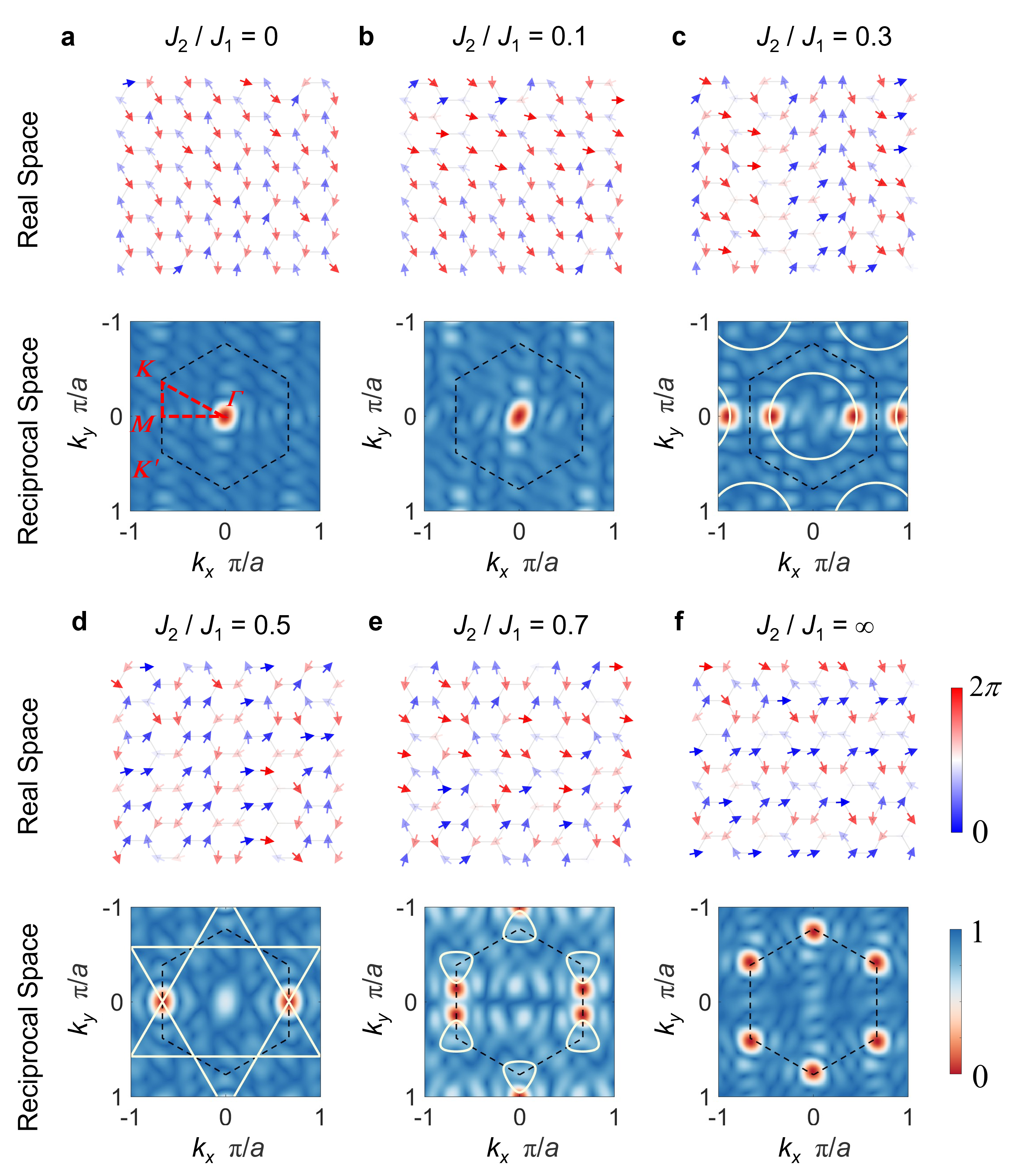}
\caption{\label{fig_s4}The experimental low-temperature states for antiferromagnetic $J_1$-$J_2$ XY model on the honeycomb lattice at $T/J=0.005$. (a)-(f) are the results for different $J_2/J_1$ ratios. For each subgraph, the first and second rows are the low-temperature states in the real space and reciprocal space, respectively. With the change of $J_2/J_1$, the peaks of spiral wave vectors in reciprocal space are located at different positions within the first Brillouin zone (1BZ). The black dashed line is the boundary of 1BZ, and the white solid line indicates the theoretical positions of the peaks.}
\end{figure}

For the antiferromagnetic $J_1$-$J_2$ XY models on the honeycomb lattice, $J_1$ (NN) and $J_2$ (NNN) interactions compete with each other, leading to geometrical frustrations and highly degenerate ground states at zero temperature \cite{di_ciolo_spiral_2014}. With our FPS, various ground states with varying the ratios of \( J_2/J_1 \) can be obtained handily (see FIG.~\ref{fig_s4}). When only NN interactions are considered, all NN spin pairs tend to align antiparallel such that spins within each sublattice are parallel alignment (FIG.~\ref{fig_s4}a). It means that the peak of the spiral wave vector (in the reciprocal space) resides at the center (\( \bm{\mathit{\Gamma}} \) point) of the 1st Brillouin zone (1BZ). This is called the Néel antiferromagnetic phase, and it will remain within the range \( 0 \leq J_2/J_1 \leq 1/6 \) (FIG.~\ref{fig_s4}a and b). When \( J_2/J_1 = 1/2 \), the competing NN and NN interactions lead to a collinear phase, where the spins in the same row parallelly align, while adjacent rows is antiparallel alignment (FIG.~\ref{fig_s4}d). In this case, the peaks of the spiral wave vector are distributed along an open path passing through the \( \boldsymbol{M} \) points. Under the limit of pure NNN interactions(\( J_2/J_1 = \infty \)), the orientation of NN spin pairs within each sublattice differs by \( 120^\circ \), and the peaks in the 1BZ are located at the \( \boldsymbol{K} \) and \( \boldsymbol{K}^\prime \) points, corresponding to a \( 120^\circ \) ordered phase (FIG.~\ref{fig_s4}f). For \( 1/6 < J_2/J_1 < 1/2 \) and \( J_2/J_1 > 1/2 \), the real-space spin configurations are void of any obvious order, but the peaks separate and are located along a closed contour centered around the \( \bm{\mathit{\Gamma}} \) point, and the \( \boldsymbol{K} \)/\( \boldsymbol{K}^\prime \) points, respectively (FIG.~\ref{fig_s4}c and e). These experimental findings conform with theoretical predictions \cite{di_ciolo_spiral_2014}, highlighting the potential of our FPS for exploring spin systems with intricate interactions.

\section{S6. The Long-Range Models}

\begin{figure}[t]
    \centering
    \includegraphics[width=0.9\linewidth]{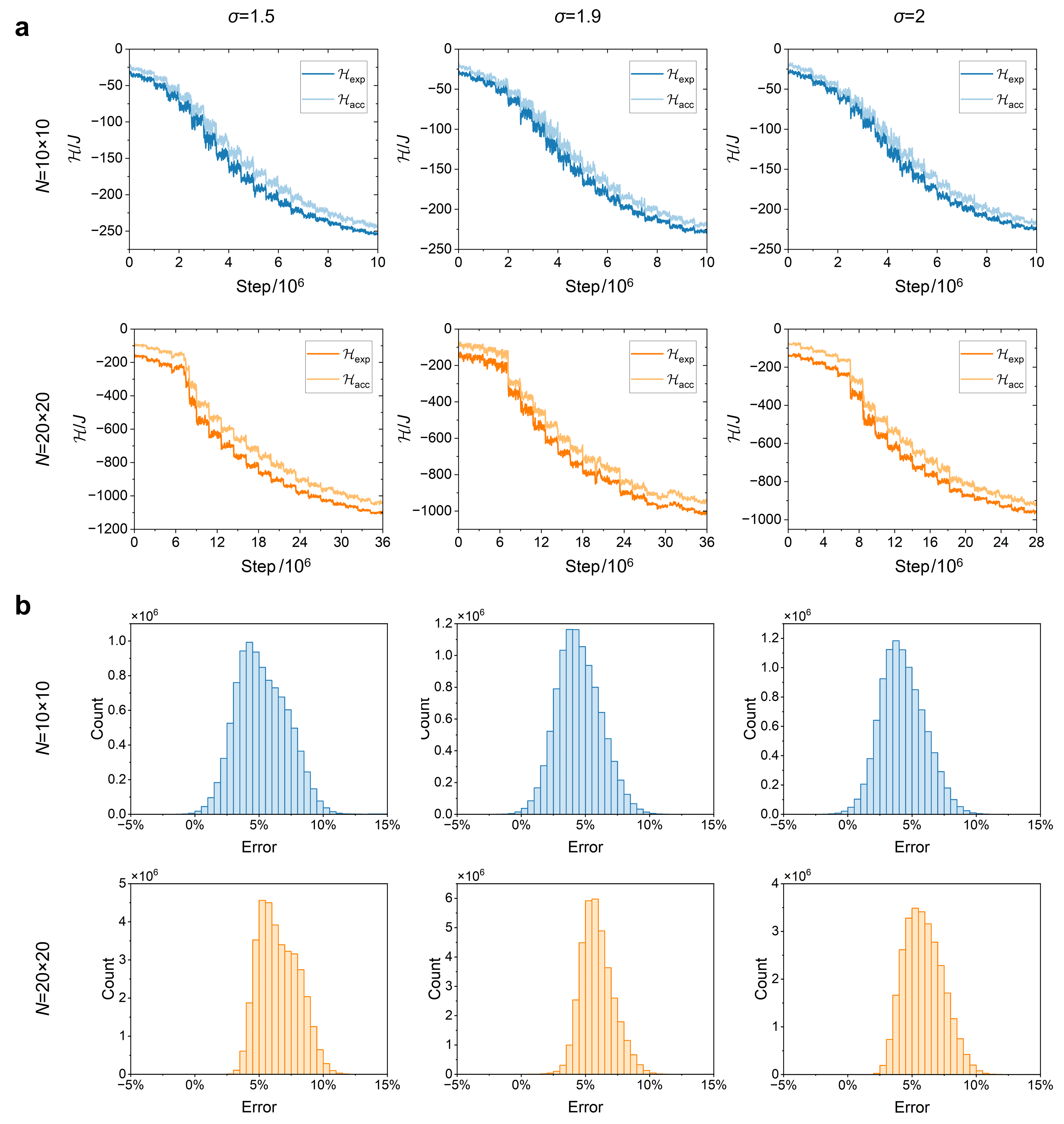}
    \caption{\label{fig_s5} Quantitative experimental results of XY model on the square lattice with LR interaction.(a) The experimentally obtained Hamiltonian $\mathcal{H}_{\mathrm{exp}}$ and corresponding theoretically accurate value $\mathcal{H}_{\mathrm{acc}}$ with different lattice sizes (blue ones for $N=10\times10$ and orange ones for $N=20\times20$) and different $\sigma$ values ($\sigma=1.5, 1.9, 2$). (b) The relative error, $\frac{\mathcal{H}_{\mathrm{exp}}-\mathcal{H}_{\mathrm{acc}}}{\mathcal{H}_{\mathrm{init}}}$, where $\mathcal{H}_{\mathrm{init}}$ is the Hamiltonian when all the spins align parallel.}
\end{figure}

Long-range (LR) interacting systems have attracted vast research interests owing to unveiling many exotic physical phenomena, where each spin interacts with all other spins. However, the calculation is intractable due to the high computational complexity with $\mathcal{O}(N^2)$. Currently, the most advanced local updating algorithm just reduces the complexity to $\mathcal{O}(N\mathrm{log}N)$ \cite{muller_fast_2023}. On the contrary, our FPS can obtain the Hamiltonian directly, regardless of the type of interaction, as long as the interaction can be synthesized through a single Fourier mask. As a result, the complexity of FPS can be significantly reduced to $\mathcal{O}(N)$. Notably, there is cluster updating algorithm also with reported complexity of $\mathcal{O}(N)$, but it is inapplicable in studies in nonequilibrium state and systems with frustration. By contrast, our method exactly reproduces the same Markov chain as the traditional Metropolis-Hastings algorithm, which can be used to study frustration.\par

In the experiment, we investigate the XY model on the square lattice with long-range, power-law decaying interactions, $J(\boldsymbol{r}_{jh})=\frac{1}{\left|\boldsymbol{r}_{jh}\right|^{(d+\sigma)}}$, where $\boldsymbol{r}_{jh}=\boldsymbol{x}_j-\boldsymbol{x}_h$ is the distance vector between the $j$-th and $h$-th spins, $d$ is the dimension of the system ($d = 2$ in our case), and $\sigma$ is the decay index characterizing the decay rate of interactions. This corresponding Hamiltonian can be described as\cite{giachetti_berezinskii-kosterlitz-thouless_2021}
\begin{equation}
        \mathcal{H}=-\sum_{(j,\,h)}J(\boldsymbol{r}_{jh})\cos\left(\theta_j-\theta_h\right)~,
\label{equ_s7}
\end{equation}
where $(j,\,h)$ denotes all spin pairs. We simulate the LR model with $\sigma=1.5,\,1.9,\,2$, all fall in the non-classical regime ($1<\sigma\leq 2$). In this regime, the LR model undergoes a second-order transition into an long-range-order phase, which is significantly different from the short-range XY model \cite{xiao_two-dimensional_2024}. The experimentally obtained Hamiltonian $\mathcal{H}_{\mathrm{exp}}$, corresponding accurate value $\mathcal{H}_{\mathrm{acc}}$, and the relative error are shown in FIG.~\ref{fig_s5}. Although there is deviation between $\mathcal{H}_{\mathrm{exp}}$ and $\mathcal{H}_{\mathrm{acc}}$, the error is almost a constant (FIG.~\ref{fig_s5}b). Because only the energy difference affects the spin flipping $\Delta\mathcal{H}$, the constant error does not impact the simulation results.

\begin{figure}[t]
\centering
\includegraphics[width=0.6\linewidth]{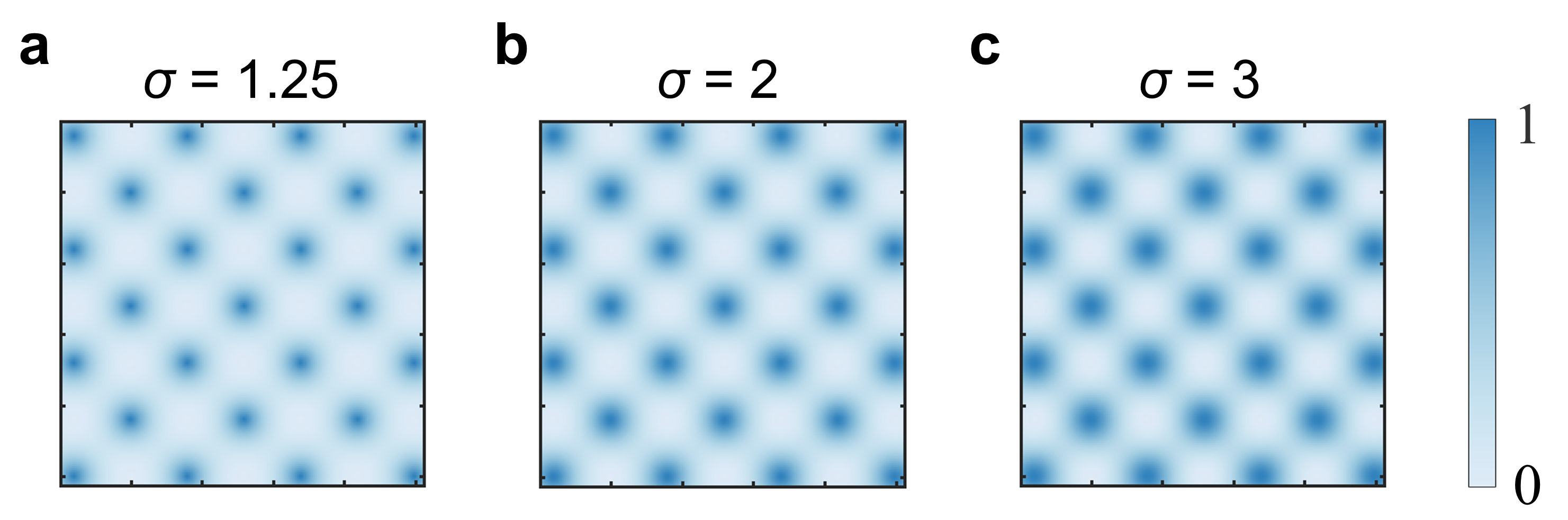}
\caption{\label{fig_r2}The Fourier masks for different long-range interactions with power-law decaying interactions. (a) $\sigma = 1.25$, (b) $\sigma = 2$, and (c) $\sigma = 3$.}
\end{figure}

\begin{figure}[t]
\centering
\includegraphics[width=0.4\linewidth]{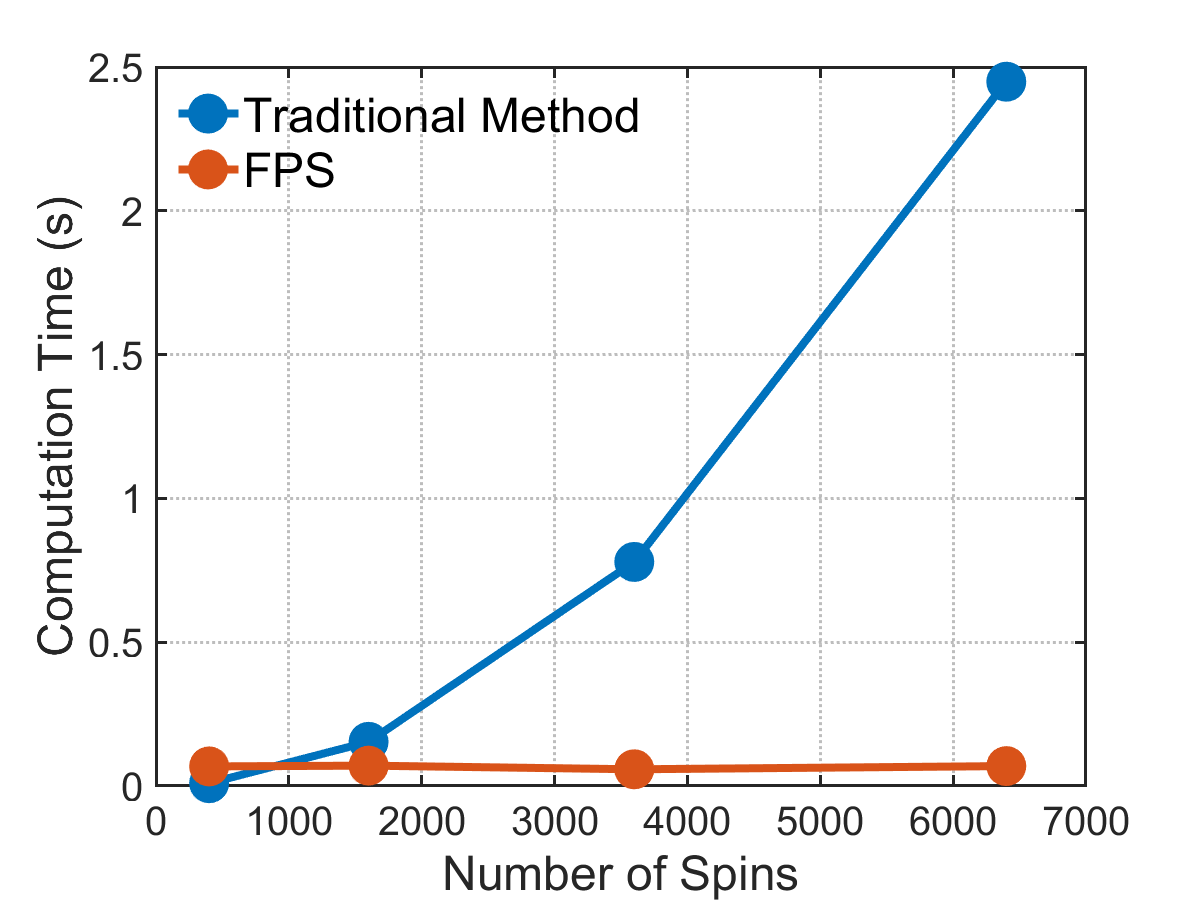}
\caption{\label{fig_r3}The time consumption of the Hamiltonian of a spin configuration with long-range interaction. The computation time of the traditional method increases with the number of spins, while that of the FPS is a constant. Each point is the result of an average of 100 different spin configurations. Here, we have adopted the ferromagnetic XY model on the square lattice with the long-range power-law decaying interactions.}
\end{figure}

It is also noted that the different types of long-range interactions require distinct Fourier masks, and longer interaction range corresponds to a smaller decay index $\sigma$ (FIG.~\ref{fig_r2}). As stated above, the time consumption of calculating different Hamiltonians are in general independent of the spin interaction range (FIG.~\ref{fig_r3}). Technically, the optical implementation of the long-range interaction Fourier masks is limited by the pixel size and modulation depth of the SLM, which can be mitigated by using a higher bit-depth SLM and a lens with longer focal length, or by switching to advanced manufactured metasurfaces, which typically offer smaller spatial sizes and higher modulation depths.

\section{S7. The Topological Invariant in FPS}
\begin{figure}[t]
\centering
\includegraphics[width=0.8\linewidth]{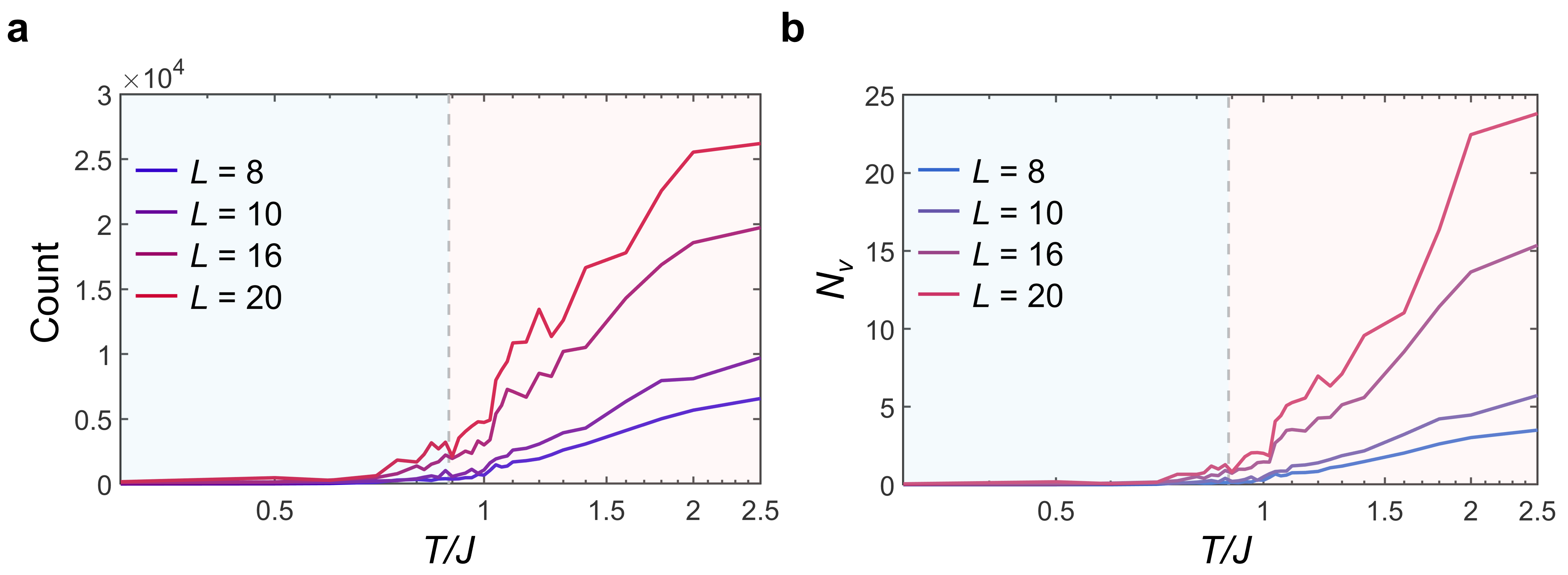}
\caption{\label{fig_r4}Topological invariant and number of unbound vortices $N_v$ at different temperatures for square lattice. (a) Counts of the nonzero topological invariants. (b) The number of unbinding vortices.}
\end{figure}

In our experiment, we used the helicity modulus to characterize the BKT phase transition. In fact, the topological invariant is defined and automatically embedded in FPS. According to the Ref. \cite{kosterlitz_nobel_2017} vortex-antivortex pairs are tightly bound below $T_\text{BKT}$, and these vortices start to bind and freely move through the lattice above $T_\text{BKT}$. In this case, the topological invariant change by 1 if a vortex-antivortex pair separates. The topological invariants are expressed as \cite{kosterlitz_nobel_2017}
\begin{equation}
\left\{ \begin{aligned}
 n_x = \frac{1}{2\pi} \int_0^L \frac{\partial \theta}{\partial x} \, dx   \\
 n_y = \frac{1}{2\pi} \int_0^L \frac{\partial \theta}{\partial y} \, dy
\end{aligned}\right.\,,
\end{equation}
which denotes the number of multiples of $2\pi$ the orientation angle $\theta$ changes. When $T<T_\text{BKT}$, the $n_x$ and $n_y$ are equal to 0 owing to no free vortex. On the contrary, when $T>T_\text{BKT}$, the free vortices start to occur, and the number of $n_x$ and $n_y$ increases with the temperature. This incurs 2D XY models possessing topologically ordered state with a finite temperature. In order to reveal the BKT phase transition, we count the number of nonzero $n_x$ and $n_y$ as a function of the temperature for all spin configurations obtained by FPS, shown in FIG.~\ref{fig_r4}a. As the results, the number of nonzero topological invariant starts to emerge and increase above the $T_\text{BKT}$, which is consistent with the theoretical prediction. \par

From the perspective of vortex-antivortex binding and unbinding, the number of free/unbinding vortices has the same behaviour with the topological invariant. In order to clear this, we also calculate the free vortex number at different temperatures, in which each vortex can be located by calculating the topological charge \cite{kosterlitz_critical_1974}
\begin{equation}
    \mathcal{C}=\frac{1}{2\pi}\oint_{\mathcal{G}}\frac{\partial \theta}{\partial s} ds
\end{equation}
where $\mathcal{G}$ is the edge of each unit cell and $\frac{\partial \theta}{\partial s}$ is the angle difference between the adjacent spins. The free vortex is defined if there is no vortex at its nearest neighbors. Finally, the results demonstrate the emergence of free vortices near the $T_\text{BKT}$ (FIG.~\ref{fig_r4}b).

$\ $

$^{*}$Y.S. and W.F. contributed equally to this work

$^{\dagger}$Corresponding author: weiru\_fan@zju.edu.cn

$^{\ddagger}$Corresponding author: dwwang@zju.edu.cn

$^{\S}$Corresponding author: hqlin@zju.edu.cn

%